\documentclass[preprint]{elsarticle}

\usepackage{amsmath}
\usepackage{amsfonts}
\usepackage{array}
\usepackage{booktabs}
\usepackage{enumitem}
\usepackage{graphicx}
\usepackage{float}
\usepackage{mathtools}
\usepackage{subcaption}
\usepackage{wrapfig}

\graphicspath{ {./images/} }
\DeclareGraphicsExtensions{.png}

\usepackage[linesnumbered,ruled,algo2e]{algorithm2e} 

\begin{document}

\begin{frontmatter}

\title{Towards an Extrinsic, CG-XFEM Approach Based on Hierarchical Enrichments for Modeling Progressive Fracture}

\author[1]{M. Keith Ballard\corref{cor1}}
\author[2]{Roman Amici}
\author[2]{Varun Shankar}
\author[3]{Lauren A. Ferguson}
\author[4]{Michael Braginsky}
\author[5]{Robert M. Kirby}

\cortext[cor1]{Corresponding author}

\address[1]{Institute for Predictive Performance Methodologies, University of Texas at Arlington Research Institute, USA}

\address[2]{Scientific Computing and Imaging Institute, University of Utah, USA}

\address[3]{Air Force Research Laboratory, Wright-Patterson Air Force Base, USA}

\address[4]{University of Dayton Research Institute, USA}

\address[5]{University of Utah, School of Computing, USA}

\begin{abstract}[Summary]{
We propose an extrinsic, continuous-Galerkin (CG), extended finite element method (XFEM) that generalizes the work of Hansbo and Hansbo to allow multiple Heaviside enrichments within a single element in a hierarchical manner. This approach enables complex, evolving XFEM surfaces in 3D that cannot be captured using existing CG-XFEM approaches. We describe an implementation of the method for 3D static elasticity with linearized strain for modeling open cracks as a salient step towards modeling progressive fracture. The implementation includes a description of the finite element model, hybrid implicit/explicit representation of enrichments, numerical integration method, and novel degree-of-freedom (DoF) enumeration algorithm. This algorithm supports an arbitrary number of enrichments within an element, while simultaneously maintaining a CG solution across elements. Additionally, our approach easily allows an implementation suitable for distributed computing systems. Enabled by the DoF enumeration algorithm, the proposed method lays the groundwork for a computational tool that efficiently models progressive fracture. To facilitate a discussion of the complex enrichment hierarchies, we develop enrichment diagrams to succinctly describe and visualize the relationships between the enrichments (and the fields they create) within an element. This also provides a unified language for discussing extrinsic XFEM methods in the literature. We compare several methods, relying on the enrichment diagrams to highlight their nuanced differences.}
\end{abstract}

\end{frontmatter}


\section{Introduction}\label{intro}

Researchers have proposed numerous computational methods for modeling the fracture of solids. Each method has its advantages for individual scenarios, but due to the complexity of fracture, no single method has won out over the others for general use.  One significant difference between approaches lies in the choice of a damage model.  Damage models treat damage as either a discrete process that localizes to a surface, or as a continuum process that affects the response of a finite volume of material. While both methods capture the far field influence of damage, discretely modeling damage yields more accurate stress fields near fracture surfaces.  In this paper, we address discrete modeling of damage, for which there are a variety of methods in the literature.  However, the genre of continuous-Galerkin (CG), extended finite element methods (XFEM) have not been capable of modeling some complex crack interactions.  This paper proposes an extension of CG-XFEM for handling such situations.

To understand where this contribution fits within the larger community of discrete damage modeling, we provide a review of the literature.  For most problems of interest to the community, the crack path is not known a priori.  Most computational methods suitable for modeling crack propagation with solution-dependent paths broadly fall into the following categories:
\begin{enumerate}[noitemsep]
\item Remeshing methods, which modify the mesh throughout the analysis to conform to inserted cracks,
\item Meshless/mesh-free methods, which only require the enrichment of a node-based mesh upon which to solve the governing equations,
\item Generalized/extended finite element methods, which account for cracks without requiring a conformal mesh by enriching the approximation functions,
\item Augmented finite element methods, which account for the displacement jump without introducing new degrees of freedom through a static condensation technique,
\item Embedded discontinuity approaches, which introduce surface elements within the volume elements to account for the discontinuity and consider the coupling between the two,
and
\item Boundary element methods, which only require a boundary mesh (including surfaces of cracks) instead of a volume mesh upon which to solve the governing equations.
\end{enumerate}

Remeshing methods combined with the classical finite element method (FEM) were some of the first methods developed for numerically predicting crack propagation.  Since the mesh must conform to the crack as it grows, one approach is to remesh the entire domain whenever cracks are modified, but this strategy is costly.\cite{Tradegard1998FEMRemeshing,Bouchard2000Crack}  Some methods effectively remesh at the element level by replacing elements cut by a crack extension with new elements, but it can be difficult to insert standard elements that conform to both the original element boundaries and the crack surface.  Polyhedral finite element methods (PFEM) extend the element library to include polyhedral elements, allowing a standard element to be cut by an arbitrary crack and replaced by two polyhedrons (or polygons in 2D).\cite{Perumal2018Brief,Wicke2007Finite,Nguyen-Hoang2017new,Teng2018adaptively,Tang2009novel}

Outside PFEM, requiring the mesh to conform to any surfaces that introduce discontinuities is very restrictive when dealing with complex geometries and large analyses.  Meshless (sometimes referred to as mesh-free) methods emerged to avoid the need for a mesh altogether, based on the early development of smoothed-particle hydrodynamics.\cite{Gingold1977SmoothedParticleHydro,Belytschko1996Meshless}  While mesh-based methods construct the trial space used to approximate the solution on a mesh, meshless methods construct the trial space only on nodes.\cite{Babuska2003Survey}  These methods are well-suited for situations where very large deformations are expected, which would result in extreme mesh distortion, or where discontinuities would move through the domain, such as moving phase boundaries or crack propagation.\cite{Belytschko1996Meshless,Nguyen2008Meshless}  Meshless methods can solve a broad class of problems that are difficult for mesh-based methods and still receive attention in the recent literature.\cite{Zhuang2012Fracture,Zhuang2011Accurate}

Babuška et al. \cite{Melenk1996partition} developed the partition of unity finite element method (PUFEM) within the genre of meshless methods to allow special functions to be used in the approximation based on a priori knowledge, building on the work of Duarte and Oden in h-p adaptive methods.\cite{Duarte1996h-p}  Borrowing concepts from meshless methods, especially PUFEM, Belytschko et al. used analytical solutions of simple fracture problems to enrich the approximation functions, while maintaining partition of unity.  This extended finite element method (XFEM) could accommodate the presence of a crack with minimal remeshing during crack propagation.\cite{Belytschko1999Elastic}  The XFEM community continued to develop methods that used local enrichment with a focus on modeling cracks and other discontinuities.\cite{Moes1999finite,BelytschkoReview}  In parallel, Babuška and collaborators at the University of Texas similarly developed the generalized finite element method (GFEM).\cite{Melenk1995On,Strouboulis2000design,Strouboulis2001generalized}  Though early developments of GFEM focused on global enrichment, the two categories of methods began to converge and were considered equivalent by Belytschko.\cite{BelytschkoReview}  Within the XFEM community, methods fall into two subgenres.  First, extrinsic methods account for enrichment by introducing additional degrees of freedom (DoFs) into the global system of equations, and the majority of XFEM variants fall into this category.\cite{BelytschkoReview,Fries2014Overview}  Second, intrinsic methods avoid introducing additional DoFs by basing the approximation functions on moving least squares functions instead of standard polynomial shape functions.\cite{Fries2014Overview,Fries2006intrinsic}

An early XFEM development especially pertinent to this work is that of Hansbo and Hansbo.\cite{Hansbo2004finite} They used a generalized Heaviside function to enrich the approximation for the presence of a crack surface, which takes the value of 1 on one side of the crack surface and 0 on the other side.  Previously, the XFEM/GFEM community proposed simply adding terms involving the Heaviside function to the approximation function and introducing additional DoFs, but this causes the original DoFs to no longer directly represent displacements.\cite{Areias2006comment}  Hansbo and Hansbo restructured the approximation function to let the original DoFs represent the displacement field on one side of the crack, and let the new DoFs represent the displacement field on the other side.  This is mathematically equivalent to the classical XFEM approach \cite{Areias2006comment}, but greatly simplifies the hierarchical view of fields that this work proposes.

Taking a different approach to extend standard FEM to account for discontinuities, the augmented finite element method (AFEM) effectively considers a remeshed element that conforms to a discontinuity and statically condenses the additional DoFs at the element level.  This results in an element-local algorithm with great computational efficiency, but introduces nonphysical discontinuities that can lead to significant error near a crack.\cite{Ling2009augmented,Naderi2016three,Jung2016Augmented}  This strategy has many advantages from the viewpoint of economy, providing algorithms that are easy to parallelize and avoiding the need to modify the global system of equations, but it also creates artifacts that can significantly degrade accuracy and requires careful estimation of the error introduced.  Highlighting and ameliorating this issue, Ma et al. recently proposed conforming-AFEM (C-AFEM), which approaches the problem as a local-global analysis and requires iterative solutions at both crack and global levels until convergence is achieved.\cite{Ma2019Conforming}  C-AFEM is a significant improvement for the AFEM community, though the nonlocal and multiscale nature of the algorithms makes parallel implementation more difficult, and the similarities to intrinsic XFEM should be noted.

In parallel to GFEM, embedded discontinuity methods were developed, accounting for the presence of a discontinuity by introducing DoFs that correspond to the response of the surface and coupling the equations with those governing the volume element.\cite{Dias-da-Costa2009comparative}  Early on, Bolzon \cite{Bolzon2001Formulation} developed a method that accounts for discontinuities for meshes that employ constant strain triangular elements.  Soon after, Alfaiate et al. \cite{Alfaiate2003Non-homogeneous} proposed two different strategies for inserting interfacial elements within a volume element.  The first strategy statically condensed the DoFs corresponding to the displacement jump, which is very similar to classical AFEM. The second strategy introduces new DoFs at the interface, which is similar to XFEM but with fewer DoFs, resulting in additional error for the displacement field.   Linder et al. \cite{Linder2007Finite} generalized Bolzon's first strategy by introducing DoFs and subsequently eliminating them from the system of equations through static condensation.  Meanwhile, Dias et al. \cite{Dias-da-Costa2009discrete} generalized the second strategy, calling it the discrete strong discontinuity approach (DSDA), by keeping the additional DoFs in the system of equations.   Finally, Dias et al. \cite{Dias-da-Costa2009Towards} developed the generalized strong discontinuity approach (GSDA) by borrowing concepts from both Linder et al. \cite{Linder2007Finite} and Dias et al.,\cite{Dias-da-Costa2009discrete} but weakly enforcing continuity of tractions between elements.  GSDA has many similarities to XFEM approaches, but the primary difference is that GSDA introduces fewer DoFs into an element to capture the presence of a discontinuity, resulting in a discontinuous displacement field across elements.

Analyses conducted by engineers for design often seek information about the region with the highest stresses, ignoring the stress state for much of the volume.  When modeling fracture, the highest stresses generally develop near material boundaries, especially crack fronts.  Consequently, for decades, researchers have been developing boundary element methods (BEM) to target such problems.  When the response is linear, this genre of methods only requires a boundary mesh, without modeling the volume, and results in a more dense system of equations compared to sparse systems encountered in FEM.  When nonlinearities are introduced, only the portion of the volume experiencing nonlinearity needs to be modeled.\cite{Aliabadi20033.02}

It should be noted that there are many other methods that do not strictly fall into the categories of methods suitable for either continuum damage models or discrete damage models.  For example, variational methods have been developed to connect the two disciplines.\cite{Sutula2018Minimum}  A genre of variational methods of particular note is phase-field models (PFM), which solve two sets of partial differential equations: one that governs the elastic response of the continuum, and one that governs the intensity of a second phase-field.  Phase-field models were originally developed to model moving weak boundaries, but researchers have applied them to fracture by treating cracks as a phase-field and minimizing the total potential energy, which is an extension of Griffith’s theory.\cite{Hakim2009Laws,Silva2013Sharp-crack,Staroselsky2019Phase}  Peridynamics is another approach outside the traditional categories, which restructures the governing equations in terms of integral quantities and thus avoids the singularities emerging in formulations that rely on partial differential equations.\cite{Silling2010Crack,agwai_predicting_2011,Ha2010Studies}

Despite the wide variety of computational methods for discretely modeling fracture, no method stands out as well-suited for problems involving the combination of complex networks of interacting cracks, nonlinear materials, and complex domains.  Additionally, many problems that fall into this category require modeling a large domain and/or with fine discretization near crack networks, making a distributed implementation crucial.  Therefore, it is greatly beneficial to develop a method that easily lends itself to a distributed implementation.

Inspired by the work of Hansbo and Hansbo \cite{Hansbo2004finite} and Iarve \cite{Iarve2010Multi-Scale,Iarve2011Mesh-independent}, this paper proposes an extrinsic, XFEM approach for modeling discontinuities that relies on a hierarchical view of Heaviside enrichments in an element, without introducing the artifacts seen in AFEM. In particular, the approach we offer:
\begin{enumerate}[noitemsep]
\item Supports an arbitrary number of Heaviside enrichments within an element, while properly accounting for the interactions between them;
\item Provides a novel DoF enumeration that naturally maintains a continuous-Galerkin (CG) solution across elements and can be quickly updated as XFEM surfaces evolve; and
\item Prioritizes element locality in all algorithms to ensure an implementation suitable for distributed computing.
\end{enumerate}

The proposed approach significantly extends the state-of-the-art for modeling complex, evolving 3D surfaces within the CG-XFEM community.  It serves as an apropos foundation for modeling progressive fracture in conjunction with crack models that smear the effect of the crack front, such as the cohesive segment method.  However, a detailed discussion of a crack model implementation is reserved for a future work.

In addition to the method itself, we also develop a lexicon using \emph{enrichment diagrams} for describing and visualizing relationships between enrichments (and the fields they create) both in a single element and between adjacent elements. This provides a unified language for discussing and comparing extrinsic XFEM approaches used for modeling fracture that rely on Heaviside enrichments. 

The next section discusses the basics of hierarchical Heaviside enrichment (HHE), which is the basis of the proposed XFEM approach.  This initial explanation focuses on defining the necessary concepts through illustrations.  In Section~\ref{diagrams}, we introduce the formal notation of enrichment diagrams, which will be used throughout the paper. In Section~\ref{hhe_implementation}, we describe the implementation of HHE, creating a new XFEM approach that is a natural extension of several existing methods in the literature.  Section~\ref{verification} provides several examples, which verify the implementation and showcase the capabilities of the method. In Section~\ref{comparison}, we use the enrichment diagram lexicon to compare several XFEM methods in the literature, highlighting key differences between them.  Finally, we conclude with a delineation of the strengths and shortcomings of our method and planned future work.

\section{Hierarchical Heaviside Enrichment (HHE)}\label{hhe}

 As discussed in the introduction, researchers have already successfully used Heaviside enrichments to model discontinuities and bi-material interfaces that do not conform to faces of elements in the mesh, and our approach extensively uses the concept of a hierarchy of enrichments in an element.  Additionally, we require that Heaviside enrichments:
\begin{enumerate}[noitemsep]
    \item Do not move once inserted,
    \item Extend to the boundary of an element or the surface of another enrichment in the same element, and
    \item Do not lie entirely along a face, edge, or node of an element.
\end{enumerate}

Since the literature does not use consistent terminology for delineating extrinsic XFEM approaches, we define terms related to a hierarchical view of Heaviside enrichments for the remainder of this section.  We borrow concepts from the method proposed in Hansbo and Hansbo \cite{Hansbo2004finite} and the phantom-node method. \cite{Song2006method}  Namely, a Heaviside enrichment introduces a second set of degrees of freedom for the element conforming to the original element support, such that each set represents the solution for a half-space on either side of the enrichment.  Let a \emph{field} refer to the solution defined by one set of DoFs in the element.  Though a field extends over the entire element, it only “physically” represents a subdomain of the element, referred to as the \emph{physical domain} of the field. A field may be similarly divided into two new fields with a second enrichment. This binary property of Heaviside enrichment (under the above assumptions) makes it natural to describe the creation of fields within an element as a full binary tree.  Each node of the tree represents a field, which can have either 0 or 2 children, and an enrichment exists where a node forks into two branches.  The term \emph{enrichment tree} refers to the binary tree view of fields and enrichments in an element.   

\begin{figure}[ht!]
    \begin{subfigure}[b]{0.45\textwidth}
      \centering
      \includegraphics[width=2in]{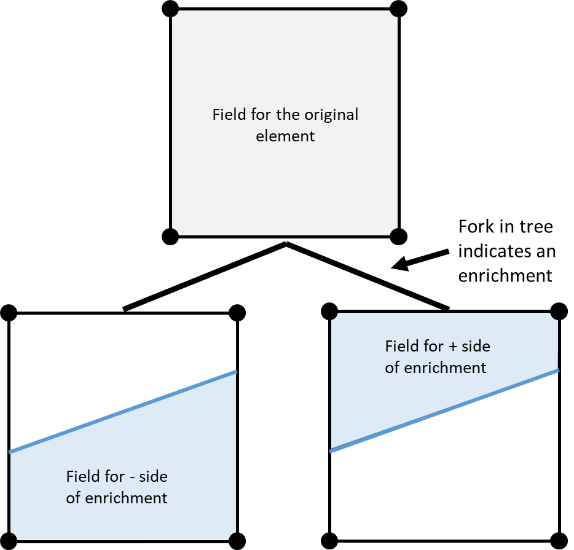}
      \subcaption{Single Heaviside enrichment}
      \label{fig:quad_enrich_a}
    \end{subfigure}
    %\hspace{.in}
    \begin{subfigure}[b]{0.45\textwidth}
      \centering
      \includegraphics[width=2.3in]{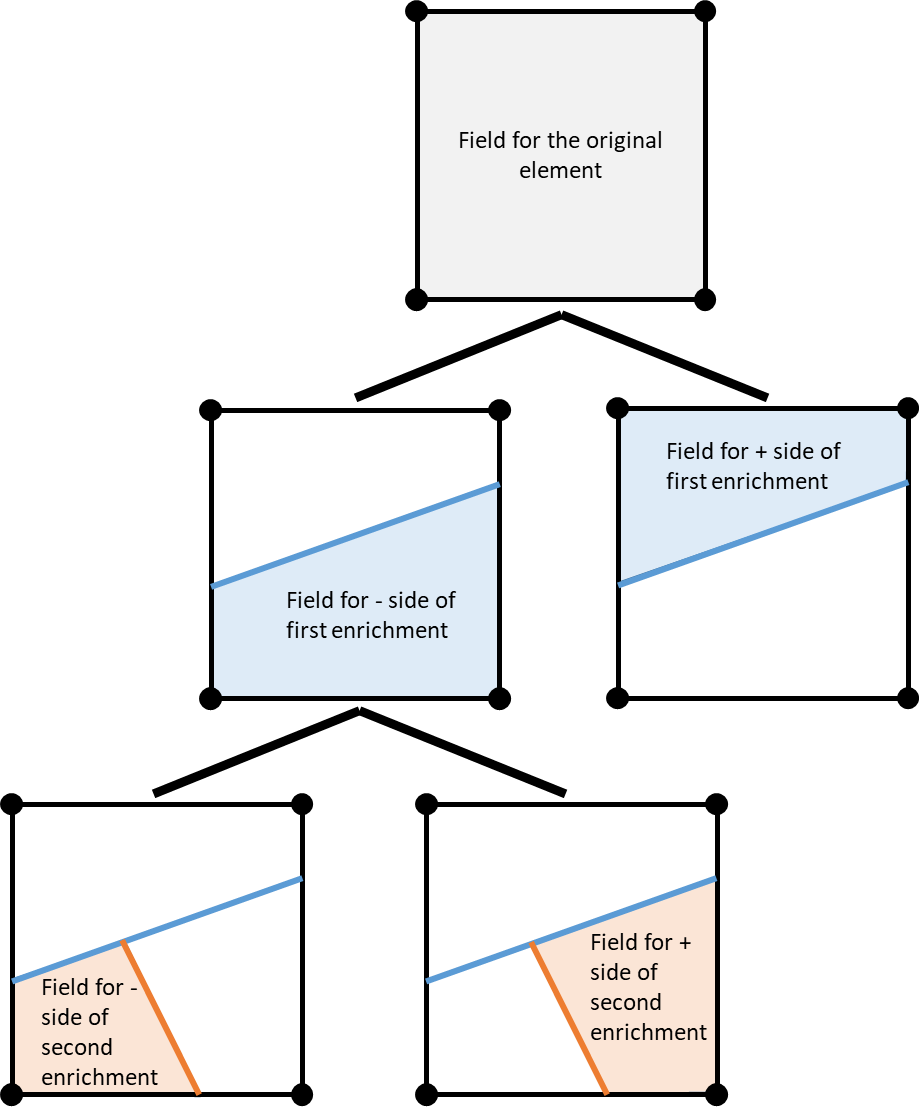}
      \subcaption{Two Heaviside enrichments}
      \label{fig:quad_enrich_b}
    \end{subfigure}
    \caption{Enrichment tree for a quadrilateral element.}
    \label{fig:quad_enrich}
\end{figure}

For the simplest case, consider a 2D quadrilateral element enriched to account for a planar discontinuity or material interface.  Figure~\ref{fig:quad_enrich_a} shows the enrichment tree.  The shaded regions in the figure indicate the physical domain of each field.  We use $+$ and $-$ to distinguish between sides of the enrichment.  The choice is arbitrary, but enforcement of interelement compatibilities requires the convention.  The enrichment trees shown in this paper will always place the field that represents the negative side of an enrichment on the left side, removing the need to label the normal vector explicitly.

The enrichment tree acknowledges the original field of the element, but only the fields that are leaves in the binary tree have degrees of freedom associated with them.  Due to this important difference, fields can be separated into two categories: \emph{basal fields}, which are the leaves in the enrichment tree, and \emph{aggregate fields}, which are interior nodes in the tree and calculated via an \emph{aggregation function} of their descendent fields, which is defined in Section~\ref{aggregation}.  In the enrichment tree, only basal fields can be enriched, and a new enrichment creates two new basal fields and changes the enriched field into an aggregate field.  If both sides of an existing enrichment should be enriched, then two enrichments must be introduced, one for each basal field on either side of the existing enrichment.  This is important for modeling phenomena like the fracture of solids, since a crack terminates at an existing crack surface or material boundary and may only ``continue'' or ``jump'' to the other side if an initiation criterion is satisfied there.  For example, Figure~\ref{fig:quad_enrich_b} shows the enrichment tree if the field on the negative side of the first enrichment in Figure~\ref{fig:quad_enrich_a} is enriched.  The resulting enrichment tree has three basal fields, two aggregate fields, and two enrichments.

In \emph{HHE}, we can hierarchically introduce Heaviside enrichments \emph{ad infinitum} within an element, each one further dividing a physical domain of a field.  Up to this point, the discussion has only considered a single element.  For multiple elements, compatibility of fields across adjacent elements requires special attention.  Let an \emph{enrichment surface} refer to a contiguous collection of element level enrichments that compose a physically meaningful surface in the global domain.  Within the context of fracture modeling, an enrichment surface could model a crack or a bi-material interface in the domain.  For CG-XFEM, the solution should be continuous throughout the global domain, except across enrichment surfaces that represent discontinuities.  The \emph{front of the enrichment surface} exists where an enrichment surface terminates interior to the global domain and is capable of growth. Thus, an enrichment in an element that terminates at another enrichment surface is not considered part of the front.  The front requires a strategy to maintain proper continuity of fields depending on what the enrichment surface models.  For example, the cohesive segment method allows a partially closed cohesive zone to exist along the surface’s front, and when the cohesive zone begins to open at the front, the cohesive surface grows.

The idea of hierarchically considering Heaviside enrichments is not novel to this work.  In the field of topology optimization, Noel et al.\cite{Noel2019Adaptive} developed an immersed boundary method using hierarchical B-splines to define Heaviside enrichments along material boundaries, including a strategy for adaptively refining the background mesh. Recently, Jahn developed an XFEM method based on hierarchical level-sets to define material boundaries for multi-phase problems, including problems with moving phase boundaries.\cite{Jahn2018automated}  Advancing the interface-enriched generalized finite element method (IGFEM) \cite{Soghrati2012}, Soghrati developed a hierarchical approach for Heaviside enrichment to ameliorate artifacts that emerge when material boundaries come close together.\cite{Soghrati2014Hierarchical}  This is not an exhaustive list of authors who have used the concept. However, to our knowledge, the hierarchical approach as presented here is a novel extension of CG-FEM for complex 3D cracking.

\section{Enrichment Diagram Notation and Terminology}\label{diagrams}

We extend the ideas and notation of the previous section to build a unified lexicon for discussing hierarchical Heaviside enrichment. With additional compatibility information, the enrichment trees become \emph{enrichment diagrams}.  The idea of using diagrams like these is not novel.  For example, Chen et al. extensively used diagrams to illustrate how the floating-node method and the phantom-node method worked, highlighting the differences between them.\cite{Chen2014EXTENED,Chen2014floating}  However, a formalized language that can efficiently describe methods using multiple enrichment surfaces is a unique contribution of this work. Table 1 defines the symbol notation for enrichment diagrams that will be used through the remainder of the paper.

\begin{table}[ht!]%
\caption{Hierarchical enrichment diagram symbols.\label{tab:diagram_notation}}%
\renewcommand{\arraystretch}{0.4}
\begin{center}
\begin{tabular}{ m{0.08\textwidth} | m{0.87\textwidth} }
\toprule
    \textbf{Symbol} & \textbf{Definition} \\
\midrule
    \includegraphics[width=0.3in]{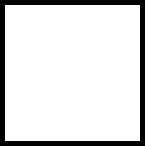} 
    &
    A box denotes a field within the enrichment tree, which can be either an \emph{aggregate field} or a \emph{basal field}.
    \\ \hline \\ 

    \includegraphics[width=0.4in]{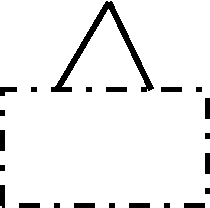}
    &
    \begin{minipage}{0.65\textwidth}
        A fork with solid lines denotes an enrichment which models a crack discontinuity that splits the parent field, resulting in:
        \begin{itemize}[noitemsep,topsep=2pt]
            \item Two new \emph{basal fields}, one for each side of the enrichment (e.g., B and C)
            \item The original basal field (e.g., A) becomes an \emph{aggregate field} after the division by the new enrichment
        \end{itemize}
        A dashed rectangle may be optionally used to refer to a specific enrichment.
    \end{minipage} \hfill%
    \begin{minipage}{0.2\textwidth}
        \includegraphics[height=0.8in]{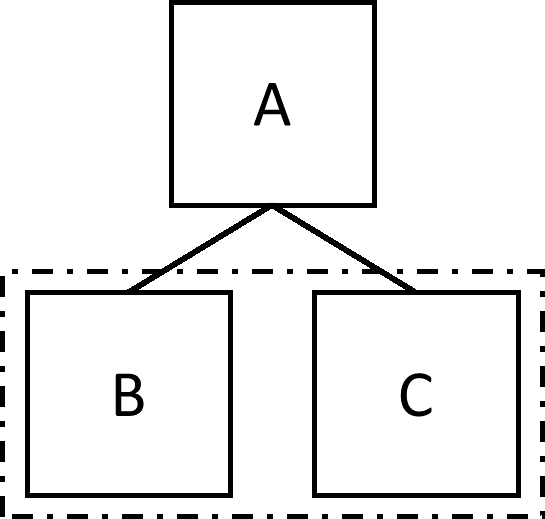}
    \end{minipage} 
    \vspace{2pt}
    \\ \hline \\

    %\begin{minipage}{\textwidth}
        \includegraphics[width=0.4in]{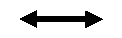}
    %\end{minipage} 
    &
    A cohesive connection between two fields (\emph{aggregate} or \emph{basal}). The fields must be in the same element or in adjacent elements.
    \vspace{2pt}
    \\ \hline \\
    
    %\begin{minipage}{\textwidth}
        \includegraphics[width=0.4in]{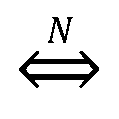}
    %\end{minipage} 
    &
    %\begin{minipage}{0.55\textwidth}
        An \emph{enriched cohesive connection} between two fields.  This connection indicates that all descendant basal fields on one side are cohesively connected to all descendent basal fields on the other side, for a total of $N$ connections.  For example, the enriched connection (left) between A and B is equivalent to the two connections (right).
    %\end{minipage} \hfill%
    %\begin{minipage}{0.3\textwidth}
        \includegraphics[height=0.8in]{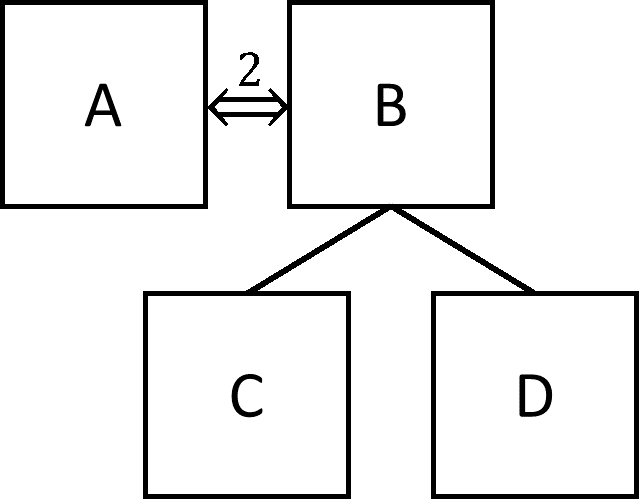}
        \hfill%
        \includegraphics[height=0.75in]{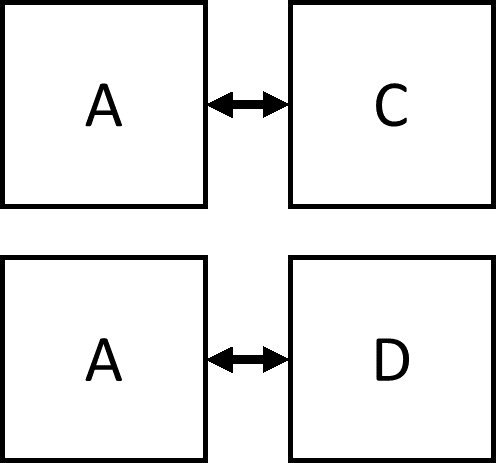}
    %\end{minipage} 
    \vspace{2pt}
    \\ \hline \\

    %\begin{minipage}{\textwidth}
        \includegraphics[width=0.4in]{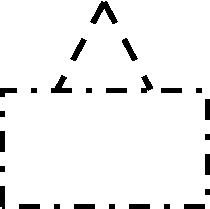}
    %\end{minipage} 
    &
    \begin{minipage}{0.65\textwidth}
        An enrichment for a bi-material interface within the element.  It is very similar to the enrichment for a crack, except that the material of each child is different (e.g. B and C).  We omit the dashed rectangle when the discussion does not refer to the enrichment.
    \end{minipage} \hfill%
    \begin{minipage}{0.2\textwidth}
        \includegraphics[height=0.85in]{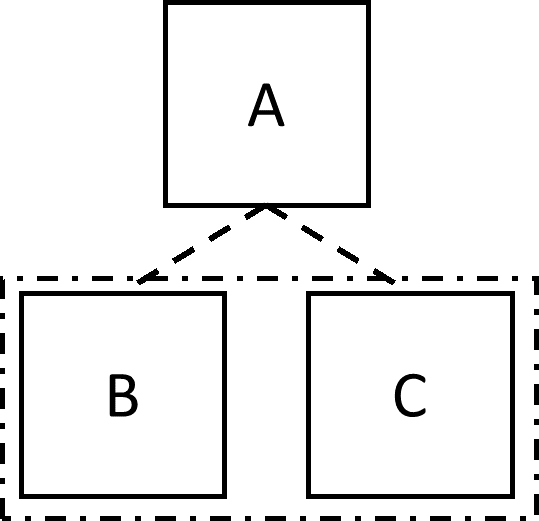}
    \end{minipage} 
    \vspace{3pt}
    \\ \hline \\

    %\begin{minipage}{\textwidth}
    \includegraphics[width=0.4in]{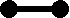}
    %\end{minipage} 
    &
    \begin{minipage}{0.5\textwidth}
        A \emph{compatibility} between two pairs of contiguous fields. The fields must be either in the same element or in adjacent elements.   For example, A is compatible with C, and B with D.
    \end{minipage} \hfill%
    \begin{minipage}{0.35\textwidth}
        \includegraphics[height=0.42in]{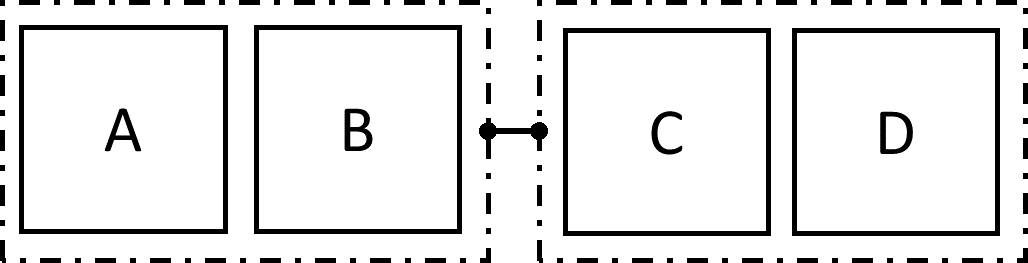}
    \end{minipage} 
    \vspace{2pt}
    \\ \hline \\

    %\begin{minipage}{\textwidth}
    \includegraphics[width=0.4in]{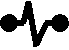}
    %\end{minipage} 
    &
    \begin{minipage}{0.5\textwidth}
        An incompatibility (or nonphysical discontinuity) between a pair of contiguous fields.  For example, A is discontinuous with C, and B is discontinuous with D.
    \end{minipage} \hfill%
    \begin{minipage}{0.35\textwidth}
        \includegraphics[height=0.42in]{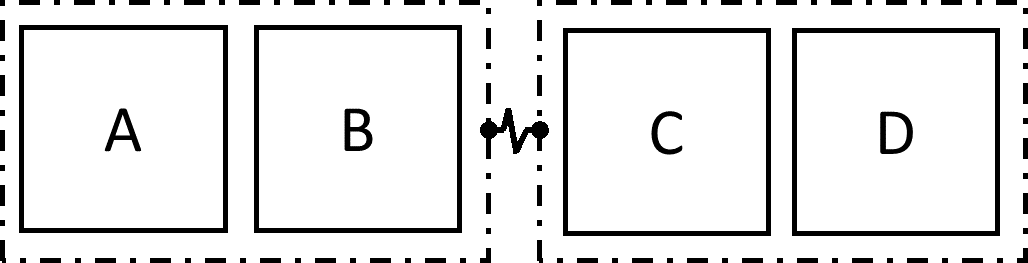}
    \end{minipage} 
    \vspace{2pt}
    \\ %\hline

\bottomrule
\end{tabular}
\end{center}
\end{table}

\clearpage

\begin{figure}[ht!]
%\setlength{\columnsep}{0.2in}
%\begin{wrapfigure}{r}{2.3in}
    \centering
    \includegraphics[width=2.2in]{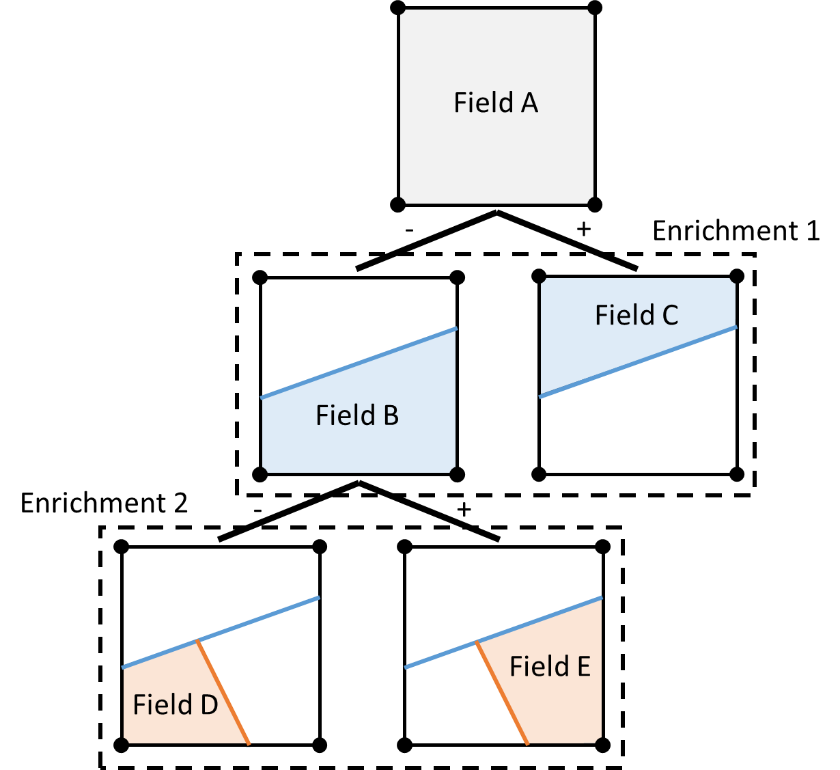}
    \caption{Enrichment diagram for notation.\label{fig:enrich_tree_term}} 
%\end{wrapfigure}
\end{figure}

The remainder of the paper relies heavily on terms that describe the relationships between fields and enrichments within the same enrichment tree.  Relationships between fields directly follow the terminology for binary trees, while relationships between enrichments are new but borrow concepts from binary trees.  To facilitate a definition of terms, Figure~\ref{fig:enrich_tree_term} shows an enrichment diagram with all fields and enrichments labeled.  Terms describing relationships in the enrichment tree are:

\begin{description}[noitemsep,parsep=0pt]
    \item[field] a function which exists over the entire element and represents the solution on its physical domain.
    \item[child field] a field that is directly below another field or created by an enrichment (e.g., Field B is a child of Field A and of Enrichment 1).
    \item[child enrichment] an enrichment directly below another field or enrichment (e.g., Enrichment 2 is a child of Field B and of Enrichment 1).
    \item[descendant fields] the set of all fields below a field or enrichment (e.g., Field B, C, D, and E are the descendant fields of Field A and of Enrichment 1).
    \item[descendant enrichments] the set of all enrichments below a field or enrichment (e.g., Enrichments 1 and 2 are the descendant enrichments of Field A; Enrichment 2 is the only descendant enrichment of Field B and of Enrichment 1).  
    \item[sibling field] one of two fields created by the same enrichment (e.g., Field B is a sibling field of Field C and vice versa). 
    \item[basal field] a field without any children, i.e., a leaf in the enrichment tree (e.g., Fields C, D, and E).
    \item[aggregate field] a field with children, i.e., an interior node in the tree (e.g., Fields A and B)
    \item[branch] one of two paths stemming from an enrichment (e.g., Enrichment 1 creates two branches, with Field B and its descendants on one branch and Field C (and any subsequent descendants) on the other).
    \item[branch sign]  the sign ($-$ or $+$) indicating which branch a field occupies for the level it exists at.  By convention, negative branches will always be on the left and positive branches will always be on the right (e.g., Fields B and D have a $-$ branch sign, while Fields C and E have a $+$ branch sign). 
    \item[parent field] the field one level up from another field or enrichment along the same branch (e.g., Field A is the parent field of Field B, Field C, and Enrichment 1).
    \item[parent enrichment] the enrichment that created a field or the enrichment one level up from an enrichment on the same branch (e.g., Enrichment 1 is the parent enrichment of Fields B and C, and of Enrichment 2).
    \item[ancestor fields] the set of all fields above a field or enrichment along the respective branch, i.e., the recursive set of parent fields (e.g., Fields A and B are ancestors of Field D, Field E, and Enrichment 2).
    \item[ancestor enrichments] the set of all enrichments above a field or enrichment along the respective branch, i.e., the recursive set of parent enrichments (e.g., Enrichment 1 is an ancestor of Fields B, C, D, and E and of Enrichment 2).
    \item[root field] a field that does not have any ancestors (e.g., Field A). 
    
\end{description}

\section{HHE Implementation}\label{hhe_implementation}

In Section~\ref{hhe}, the concepts of hierarchical Heaviside enrichment (HHE) were discussed at a high level. This section describes our implementation of the HHE theory in an XFEM formulation for infinitesimal strain elasticity in the context of fracture. To simplify the discussion in this section, enrichment surfaces will represent cracks.  Additionally, as previously stated, we impose that an enrichment is not allowed to lie along an edge or face of an element nor allowed to exactly intersect a node of an element.  When a situation arises that would violate this restriction, we move the enrichment slightly to avoid coincidence with the face, edge, or node.  This restriction ensures that the enrichment tree remains a full binary tree. Note that this paper focuses on the pieces of the analysis framework needed to account for discontinuities in the solution field. Therefore, the implementation of a crack model will be left to a future publication.

\subsection{Aggregation of Fields}\label{aggregation}

As discussed in Section~\ref{hhe}, degrees of freedom (DoFs) correspond to basal fields, and an aggregate field is a function of its descendant basal fields.  This section delineates the aggregation function for elasticity for a single enrichment and then extends to the general case of multiple enrichments. For elasticity, DoFs correspond to displacements at the nodal positions.  Before enrichment, the displacement field at any location within the element, $\bar{u}(\bar{x})$, is given by
\begin{align}
\bar{u}\left(\bar{x}\right)=\sum_{i=1}^{N}{\phi_i\left(\bar{x}\right){\bar{u}}_i},
\end{align}
where overbar denotes a vector quantity, ${\bar{u}}_i$ denotes the displacement at node $i$, $\phi_i\left(\bar{x}\right)$ denotes the element shape function which has value 1 at node $i$, and $N$ is the number of nodes in the element.  After Heaviside enrichment, the element is divided into two half-spaces, each with its own basal field to represent displacement on its side of the enrichment. Let ${\bar{u}}^-\left(\bar{x}\right)$ refer to the displacement field on the negative side of the enrichment and ${\bar{u}}^+\left(\bar{x}\right)$ refer to the displacement field on the positive side of the enrichment, with these signs assigned based on a vector normal to the enrichment surface.
The displacement fields are calculated from the respective nodal values using the existing element shape functions via
\begin{align}
{\bar{u}}^-\left(\bar{x}\right) &= \sum_{i=1}^{N}{\phi_i(\bar{x})\ {\bar{u}}_i^-}, \; \mathrm{and} \\
{\bar{u}}^+\left(\bar{x}\right) &= \sum_{i=1}^{N}{\phi_i(\bar{x})\ {\bar{u}}_i^+}.
\end{align}

Then, the \emph{aggregation function} yields the aggregate displacement field, $\bar{u}\left(\bar{x}\right)$, by
\begin{align}
\bar{u}\left(\bar{x}\right)=H\left(\bar{x}\right){\bar{u}}^+\left(\bar{x}\right)+\left(1-H\left(\bar{x}\right)\right){\bar{u}}^-\left(\bar{x}\right),
\end{align}
where $H\left(\bar{x}\right)$ is the generalized Heaviside function that takes the value of 0 on the negative side of the enrichment and 1 on the positive side.  Effectively, a parent field is a convex combination of its two child fields.  If a child field itself is an aggregate field, then the same equation recursively applies down the tree.

For the remainder of this paper, let lowercase Greek letters denote fields, uppercase script letters denote enrichments, and double stroke letters denote sets of fields or enrichments.  Furthermore, let superscripts denote to which field or enrichment a function belongs.  Finally, let subscripts refer to the containing element.

In the general case of multiple enrichments, it is useful to have an expression for $\bar{u}\left(\bar{x}\right)$ that does not rely on recursion, which is given by
\begin{align}
\bar{u}\left(\bar{x}\right)=\sum_{\beta\in\mathbb{B}_e}{C^\beta\left(\bar{x}\right){\bar{u}}^\beta\left(\bar{x}\right)},
\end{align}
where $\mathbb{B}_e$ is the set of basal fields in element $e$, and the coefficient $C^\beta\left(\bar{x}\right)$ is a function of $\beta$ and all of $\beta$’s ancestor fields, denoted by $\mathbb{A}_e^\beta$.  With $\mathcal{A}$ as the parent enrichment of $\alpha$, and $\mathcal{B}$ as the parent enrichment of $\beta$,  $C^\beta\left(\bar{x}\right)$ is given by
\begin{align}
C^\beta\left(\bar{x}\right) &= \prod_{\alpha\in\mathbb{A}_e^\beta\cup\left\{\beta\right\}}{h^\alpha\left(\bar{x}\right)},\ \mathrm{where}\\
h^\alpha\left(\bar{x}\right) &= \begin{cases}
1-H^\mathcal{A}\left(\bar{x}\right), & \mbox{ if } \alpha \mbox{ is on the  negative side of } \mathcal{A} \mbox{,} \\
H^\mathcal{A}\left(\bar{x}\right), & \mbox{ if } \alpha \mbox{ is on the positive side of } \mathcal{A} \mbox{,} \\
1, & \mbox{ if } \alpha \mbox{ is the root field}.
\end{cases}
\end{align}

\begin{figure}[th!]
\centering
\includegraphics[height=2.2in]{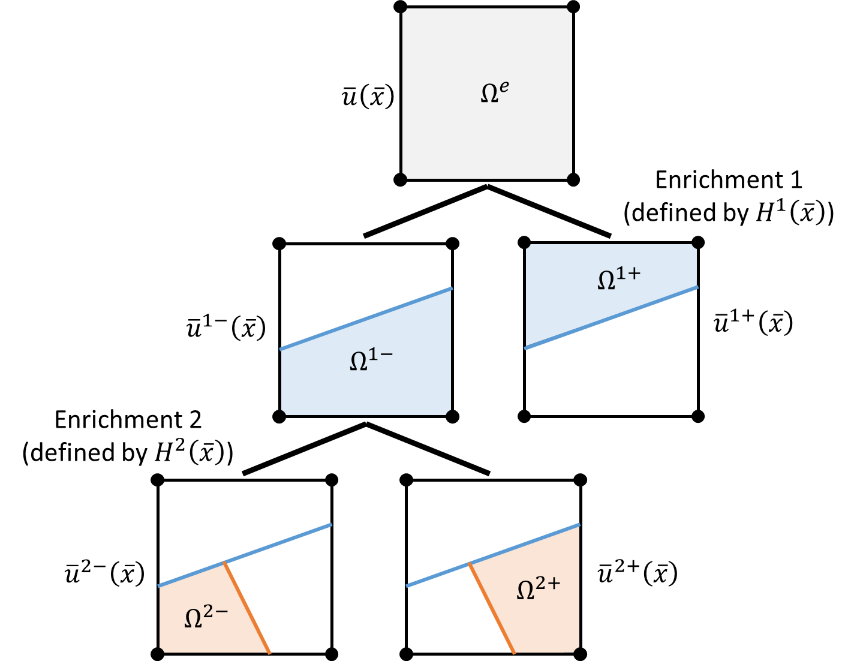}
\caption{Enrichment diagram from Figure~\ref{fig:quad_enrich_b} with labeled fields and enrichments.\label{fig:quad_enrich_lab}}
\end{figure}

Figure~\ref{fig:quad_enrich_lab} shows the application of these definitions to the situation in Figure~\ref{fig:quad_enrich_b}, where ${\bar{u}}^-\left(\bar{x}\right)$ was enriched.  The basal field indices for this example are $\mathbb{B}_e=\left\{1\mathrm{+},2\mathrm{-},2\mathrm{+}\right\}$; recall ${\bar{u}}^{1-}\left(\bar{x}\right)$ is an aggregate field after enrichment.  The root aggregate displacement field, $\bar{u}\left(\bar{x}\right)$, is given by
\begin{align}
\bar{u}\left(\bar{x}\right) &= C^{1+}\left(\bar{x}\right){\bar{u}}^{1+}\left(\bar{x}\right)+C^{2-}\left(\bar{x}\right){\bar{u}}^{2-}\left(\bar{x}\right)+C^{2+}\left(\bar{x}\right){\bar{u}}^{2+}\left(\bar{x}\right),\ \mathrm{where} \\
C^{1+}\left(\bar{x}\right) &= H^1\left(\bar{x}\right), \\
C^{2-}\left(\bar{x}\right) &= \left(1-H^1\left(\bar{x}\right)\right)\left(1-H^2\left(\bar{x}\right)\right),\ \mathrm{and} \\
C^{2+}\left(\bar{x}\right) &= \left(1-H^1\left(\bar{x}\right)\right)H^2\left(\bar{x}\right).
\end{align}

Recall that a basal field can be evaluated at any location within the element, but the function $h^\alpha$ is only nonzero over the physical domain of the field, thus the coefficient $C^\beta\left(\bar{x}\right)$ is only nonzero in a subset of the domain of the element, $\Omega_e$. Note then that an aggregate field $\beta$ also has a physical domain. In general, the physical domain of any field $\beta$, $\Omega_e^\beta$, is given by
\begin{align}
\Omega_e^\beta\subseteq\Omega_e \; | \;  \forall \, \bar{x} \in\Omega_e\rightarrow\ C^\beta\left(\bar{x}\right)=1.
\end{align}

For the situation depicted in Figure~\ref{fig:quad_enrich_lab}, the physical domain of each field is shaded.  For integrals over the element involving $\bar{u}\left(\bar{x}\right)$ or a quantity derived from $\bar{u}\left(\bar{x}\right)$, it is convenient to integrate piecewise, since
\begin{align}
\bar{u}\left(\bar{x}\right)={\bar{u}}^\beta\left(\bar{x}\right), \;  \forall \, \bar{x}\in\Omega_e^\beta, \; \beta\in\mathbb{B}_e. \label{eq:aggregate_u}
\end{align}

\subsection{Finite Element Model with Enriched Elements}\label{finite_element_model}

The implementation and theory up to this point hold for any type of physics for which the solution field contains discontinuities.  This section develops HHE for linearized infinitesimal strain theory.  The weak form of the governing equation for infinitesimal strain elasticity is provided in many textbooks \cite{Sadd2014Chapter}.  Without body forces, the weak form becomes
\begin{align}
\int{\widetilde{\sigma}\left(\bar{x}\right) : \delta\widetilde{\varepsilon}\left(\bar{x}\right) d\Omega}=\int{\bar{t}\left(\bar{x}\right)\cdot\partial\bar{u}\left(\bar{x}\right)d\Gamma}, \label{eq:weak_form}
\end{align}
where $\widetilde{\sigma}\left(\bar{x}\right)$ is the rank-2 stress tensor, $\widetilde{\varepsilon}\left(\bar{x}\right)$ is the rank-2 strain tensor, and $\bar{t}\left(\bar{x}\right)$ is the traction vector.  For a linear elastic material, stress is linearly related to strain through a rank-4 tensor that represents the constitutive relation, $\widetilde{C}$, so Eqn.~\ref{eq:weak_form} can be rewritten as
\begin{align}
\int{C_{kl}\left(\bar{x}\right)\varepsilon_l\left(\bar{x}\right)\delta\varepsilon_k\left(\bar{x}\right)\ d\Omega}=\int{t_j\left(\bar{x}\right)\delta u_j\left(\bar{x}\right)d\Gamma}, \label{eq:weak_form2}
\end{align}
where stress, strain, and the constitutive relation use Voigt notation.  This holds for any arbitrary volume, including for a single element, $\Omega_e$.  However, the form used for classical FEM requires modification for enriched elements.  If $u_j^\beta$, $v_j^\beta$, and $w_j^\beta$ denote the $x$, $y$, and $z$ components of ${\bar{u}}^\beta\left(\bar{x}\right)$, respectively, evaluated at node $j$ within an element, then all the nodal values of field $\beta$ in an enriched element can be arranged as the vector 
\begin{align}
\mathbf{q}^\beta &= \left[u_1^\beta,\ v_1^\beta,\ w_1^\beta,\ \ldots,u_N^\beta,\ v_N^\beta,\ w_N^\beta\ \right] \; \forall \, \beta\in\mathbb{B}_e,
\end{align}
where $N$ is the number of nodes in the element and $\mathbb{B}_e$ is the set of basal fields in the element.  For a given basal field within the element, ${\bar{u}}^\beta\left(\bar{x}\right)$ is a function of the nodal displacement vector, $\mathbf{q}^\beta$, and a matrix of shape functions, $\mathbf{A}$, where
\begin{align}
u_i^\beta\left(\bar{x}\right) &= A_{ij}\left(\bar{x}\right)q_j^\beta,\ \ \mathrm{and}\\
\mathbf{A} &= \left[\begin{matrix}\phi_1\left(\bar{x}\right)&0&0&\ldots&\phi_N\left(\bar{x}\right)&0&0\\0&\phi_1\left(\bar{x}\right)&0&\ldots&0&\phi_N\left(\bar{x}\right)&0\\0&0&\phi_1\left(\bar{x}\right)&\ldots&0&0&\phi_N\left(\bar{x}\right)\\\end{matrix}\right].
\end{align}

The matrix $\mathbf{A}$ involves the same shape functions for all fields within an enriched element.  Using this with Eqn.~\ref{eq:aggregate_u} yields the root aggregate displacement field, $\bar{u}\left(\bar{x}\right)$, in indicial notation
\begin{align}
u_i\left(\bar{x}\right)=A_{ij}\left(\bar{x}\right)q_j^\beta, \;  \forall \, \bar{x}\in\Omega_e^\beta,\ \beta\in\mathbb{B}_e.
\end{align}

Typically, the matrix $\mathbf{B}$ denotes the partial derivative of the strain with respect to the nodal displacement vector, $\mathbf{q}^\beta$.  Like $\mathbf{A}$, the form of $\mathbf{B}$ is the same for both HHE and classical FEM, and it is only a function of the derivatives of the shape functions for the element.  With this, the strain becomes
\begin{align}
\tilde{\varepsilon}\left(\bar{x}\right)=\mathbf{B}\left(\bar{x}\right)\mathbf{q}^\beta,\; \bar{x}\in\Omega_e^\beta,\ \beta\in\mathbb{B}_e.
\end{align}

Since the displacement and strain within the element are piecewise defined, the variational form of the governing equation must be piecewise integrated.  Let $\mathbb{B}_e$ denote the set of basal fields and $\Gamma_e^\beta$ be the boundary of $\Omega_e^\beta$.  Using piecewise integration and rearranging to separate the variation of the nodal displacements, the variational form of the governing equation is
\begin{equation}
\begin{gathered}[b]
\sum_{\beta\in\mathbb{B}_e}{\left(\int{B_{ki}\left(\bar{x}\right)C_{kl}\left(\bar{x}\right)B_{lj}\left(\bar{x}\right)d\Omega_e^\beta}q_j^\beta - \int{t_m\left(\bar{x}\right)A_{mi}\left(\bar{x}\right)d\Gamma_e^\beta}\right)\delta q_i^\beta} \\ =0,\; \forall\, \beta\in\mathbb{B}_e, \label{eq:var_form}
\end{gathered}
\end{equation}
but $\delta\mathbf{q}^\beta$ is arbitrary and independent of $\delta\mathbf{q}^\gamma\;\forall\,\gamma\in\mathbb{B}_e\setminus\left\{\beta\right\}$.  Therefore, the expression within the parentheses must be zero for each basal field $\beta$.  Notice that the finite element model forms a block system of equations of size $\left|\mathbb{B}_e\right|$.

Since enrichments introduce new boundaries into the domain, it is useful to distinguish between the new boundaries that intersect the boundary of the element and those that are internal to the element for each basal field, $\beta$.  Let $\Gamma_{ext}^\beta$ denote the external boundary of $\beta$, and let $\Gamma_{int}^\beta$ denote the internal boundary, i.e.,
\begin{align}
\Gamma_{ext}^\beta &= \Gamma_e^\beta\cap\Gamma_e,\; \forall \, \beta\in\mathbb{B}_e,\ \mathrm{and}\\
\Gamma_{int}^\beta &= \sum_{\gamma\in\mathbb{B}_e \setminus \left\{\beta\right\}}{\Gamma_e^\beta\cap\Gamma_e^\gamma},\; \forall  \, \beta\in\mathbb{B}_e.
\end{align}

Note that the intersection of three or more basal field boundaries is either an empty set or a single point.  Since $\Gamma_e^\beta=\Gamma_{int}^\beta\cup\Gamma_{ext}^\beta$, the finite element model in Eqn.~\ref{eq:var_form} can be rearranged as
\begin{equation}
\begin{gathered}[b]
\int{B_{ki}\left(\bar{x}\right)C_{kl}\left(\bar{x}\right)B_{lj}\left(\bar{x}\right)d\Omega_e^\beta} q_j^\beta \, - \sum_{\gamma\in\mathbb{B}_e \setminus \left\{\beta\right\}}{\int{t_j\left(\bar{x}\right)A_{ji}\left(\bar{x}\right)d\left(\Gamma_e^\beta\cap\Gamma_e^\gamma\right)}} \\
\,- \int{t_j\left(\bar{x}\right)A_{ji}\left(\bar{x}\right)d\Gamma_{ext}^\beta}=0, \;\forall\,\beta \in \mathbb{B}_e. \label{eq:hhe_fea_equation}
\end{gathered}
\end{equation}

The sum of the integrals over the intersections of basal field boundaries is important for a crack model and should receive special attention.  Section~\ref{diagrams} used the term \emph{enriched cohesive zone} at a conceptual level. Mathematically, an enriched cohesive zone ensures that the sum of the integrals over the intersection of basal field boundaries for all combinations of basal fields is accurately calculated.  For this paper, all enrichments are traction-free, so the integrals over the internal boundaries are all zero.  

Under this assumption, the finite element model is rearranged to form the system of linear equations 
\begin{align}
\mathbf{K}_e\mathbf{q}_e=\mathbf{F}_e.
\end{align}  
The block stiffness matrix, $\mathbf{K}_e$, for the element becomes block diagonal, with each diagonal block, $\mathbf{K}_e^\beta$, defined by
\begin{align}
\mathbf{K}_\mathbf{e}^\beta=\int{\mathbf{B}^\mathsf{T}\left(\bar{x}\right)\mathbf{C}\left(\bar{x}\right)\mathbf{B}\left(\bar{x}\right)d\Omega_e^\beta.} \label{eq:hhe_Kmatrix}
\end{align}
The off-diagonal blocks are nonzero only when tractions across internal boundaries are nonzero.  Additionally, the force vector, $\mathbf{F}_e$, may be defined via blocks $\mathbf{F}_\mathbf{e}^\beta$ for each basal field $\beta$, given by
\begin{align}
\mathbf{F}_\mathbf{e}^\beta=\int{\mathbf{A}^\mathsf{T}\left(\bar{x}\right)\bar{t}\left(\bar{x}\right)d\Gamma_{ext}^\beta}.
\end{align}
Closed-form solutions for these integrals are generally difficult to obtain, so we evaluate the expressions using numerical integration, as described in Section~\ref{integration}.

\subsection{Enrichment Surface Discretization and Growth}\label{representation}

We require a representation of enrichment surfaces to track where the surface lies and define boundaries of the physical domains of the basal fields.  A representation of an enrichment surface can be either implicit (defined by a function evaluated on the original mesh), or explicit (composed of a collection of surface elements).  Implicit representations benefit from the many established methods for evolving level-sets, easing the implementation of growth algorithms.  On the other hand, explicit representations are convenient for modeling complex behavior across enrichment surfaces, such as a cohesive law governing the opening of cracks or oxygen diffusion through a crack network, since explicit representations provide a mesh for the required calculations.  We implement both representations, using the implicit representation to determine how an enrichment surface evolves and the explicit representation for numerical integration and topological checks.

\subsubsection{Level-Set Representation}

We use the signed distances evaluated at each node of an element to implicitly represent where an enrichment lies within an element via the level-set method.  We interpolate the nodal signed distance values to any point within the element using a set of basis functions and consider the enrichment to lie along the 0-isosurface.  We restrict the interpolation of the signed distance to linear polynomials, regardless of the order of the element, in order to avoid multiple intersections along a single edge.  Additionally, we require a signed distance value at a nodal position to be non-zero, avoiding the possibility of a level-set exactly going through a node, edge, or face and guaranteeing that the level-set will cut through some volume within the element.  If a signed distance value of zero occurs at a nodal position, it is slightly perturbed.

To ease a distributed implementation, we give a strong emphasis to the locality of algorithms within this framework.  Consequently, we selected the element-local level-set method proposed by Duan et al. \cite{Duan2009Element-local}.  While evolving level-sets, Duan et al. recognized that there is a competition between minimizing the error between the represented crack surface and the surface predicted by the fracture model and minimizing the discontinuity between an extension of the crack front with the existing crack surface. Consequently, they devised a method to find a level-set within an element through least-squares minimization of these two competing error terms.  The method relies on the fact that the set of nodal signed distance values is unique to each element, which means that two adjacent elements can have different signed distance values at their shared nodes.  A weighting parameter is introduced to determine the relative importance of each of the two errors, and the authors recommended a value of 1 based on a parametric study.  This type of level-set method does not enforce $C_0$ continuity of the surface across adjacent elements that it intersects, but it does require that the sign of the nodal signed distances match across adjacent elements.

\subsubsection{Polygonal Discretization}

After an implicit representation is created for a new enrichment, we create the explicit representation based on the intersection of the level-set with the physical domain of the parent field, which is approximated by a set of bounding polygons forming a polyhedron.  Since the explicit representation is used for volume integrals over the physical domain of a basal field, the polygons are expressed in the parametric coordinate system of the element, avoiding the need to map from the global coordinate system to the parametric coordinate system of the element during numerical integration.  Additionally, we do not require that the polygons be planar, but they must be a trilinear function in the element's parametric coordinate system.  Although the implicit representation of an enrichment extends throughout the entire element, the explicit representation is bound by the physical domain of the parent field.  Furthermore, when a new enrichment is introduced, all the intersected polygons that bound the basal field being enriched are cut (or divided) by the new enrichment.

\begin{figure}[ht!]
\centering
    \begin{subfigure}[t]{0.3\textwidth}
        \centering
        \includegraphics[height=1.5in]{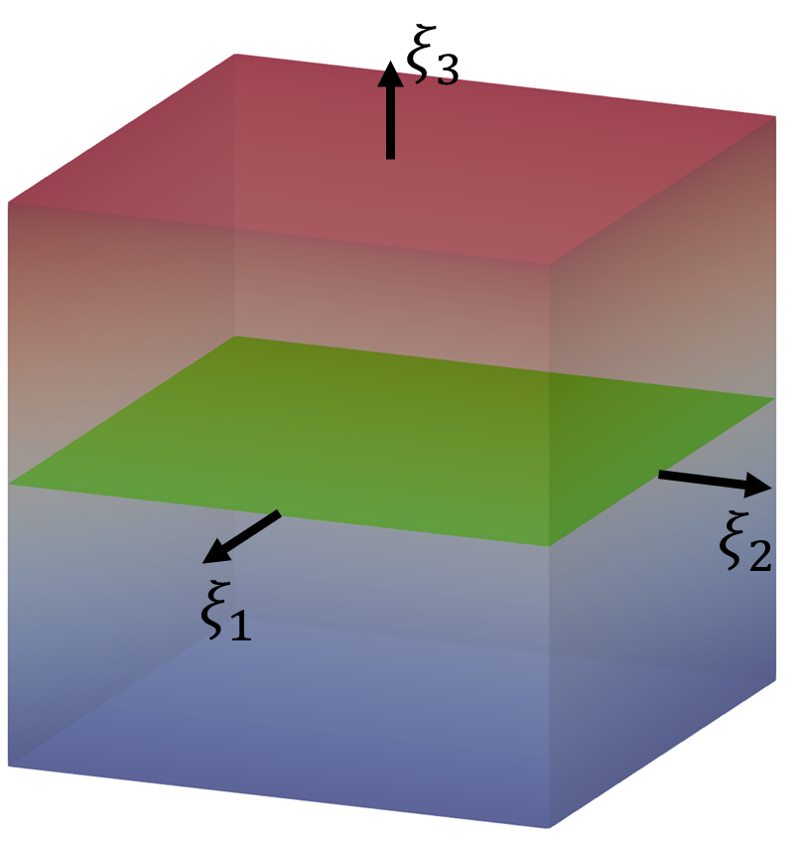}
        \caption{Implicit representation of enrichment 1 defined by $f^1(\bar{\xi})=\xi_3$\label{fig:hex_rep_a}}
    \end{subfigure} %
    ~ 
    \begin{subfigure}[t]{0.3\textwidth}
        \centering
        \includegraphics[height=1.5in]{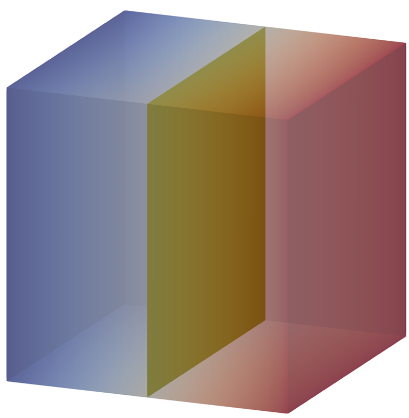}
        \caption{Implicit representation of enrichment 2 defined by $f^1(\bar{\xi})=\xi_2$\label{fig:hex_rep_b}}
    \end{subfigure} %
    ~ 
    \begin{subfigure}[t]{0.3\textwidth}
        \centering
        \includegraphics[height=1.5in]{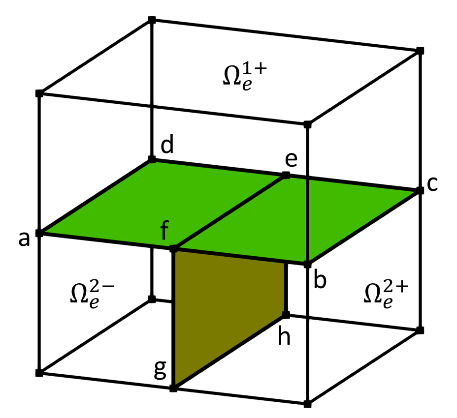}
        \caption{Explicit representation of both enrichments\label{fig:hex_rep_c}}
    \end{subfigure}
\caption{Implicit and explicit representations of two enrichments in a 8-node hexahedron element.\label{fig:hex_rep}}
\end{figure}

For example, consider an 8-node hexahedron element with a first enrichment defined by a signed distance function $f^1(\bar{\xi})=\xi_3$, which is the plane $\xi_3 = 0$, and a second enrichment defined by the signed distance function $f^2(\bar{\xi})=\xi_2$ (the plane $\xi_2=0$). Figures~\ref{fig:hex_rep_a}~and~\ref{fig:hex_rep_b} show the implicit representations for both enrichments, respectively, while Figure~\ref{fig:hex_rep_c} shows the explicit representations for both enrichments.  Note that the first enrichment restricts the explicit representation of the second enrichment. The second enrichment also caused the first enrichment to be divided into two polygons with vertices $\left\{a,f,e,d\right\}$ and $\left\{f,b,c,e\right\}$, respectively.  

The polygonal discretization of enrichments (and faces) of an element allows fast topological queries that are required for growing enrichment surfaces and finding compatibilities for basal fields and enrichments, which we discuss in the next section.  For each polygon, we store to which enrichment or face it represents, and for each edge of the polygon, we store to which element edge it lies along (if any).  This information efficiently enables the following queries:
\begin{enumerate}[noitemsep,align=left, leftmargin=3.5em]
    \item[\textbf{Query 1.}] \label{itm:query_1} Whether the level-set for an enrichment in one element intersects the physical domain of a basal field in an adjacent element through the shared topology of the two respective elements,
    \item[\textbf{Query 2.}] \label{itm:query_2} Whether the level-set for an enrichment in one element intersects the shared topology between its parent element and an adjacent element, and 
    \item[\textbf{Query 3.}] \label{itm:query_3} Whether the physical domain of a basal field touches the shared topology between its parent element and an adjacent element.
\end{enumerate}

Query 1 is critical to the algorithms described later in this paper.  Namely, it is needed when testing if a given enrichment may grow into a given basal field and when determining if a new enrichment in one element should be compatible with any existing enrichments in adjacent elements.  On the other hand, Queries 2 and 3 serve as filters to short-circuit complex logic involving enrichment tree traversals for cases when the topological predicate can guarantee that a more expensive enrichment tree predicate will not be met.  Importantly, storing this topology information alongside the polygonal discretization is not required to perform any of the three queries mentioned, since they can be based entirely on the element topology and the signed distance values of all enrichments in relevant elements.  However, storing this information significantly reduces the computational time required (at the cost of more memory).  Thorough profiling is required to understand the details of this trade-off.

\subsubsection{Growth Algorithm}\label{growth_algorithm}

To advance the front of an enrichment surface, we must first determine the basal fields into which it is topologically permissible for the front to grow.  This determination involves \emph{compatibilities} between enrichments, which must be distinguished from \emph{compatibilities} between basal fields, which was defined earlier in Table~\ref{tab:diagram_notation}.  Let a \emph{compatibility} between an enrichment in one element and an enrichment in an adjacent element mean that the solution jump should be $C^0$ continuous along the two enrichments across the adjacent elements.  Given this definition, the conditions that must be satisfied to permit an enrichment $\mathcal{A}$ that lies along the front to advance into a basal field $\beta$ are as follows:

\begin{enumerate}[noitemsep,align=left, leftmargin=3.5em]
    \item[\textbf{Condition 1.}] The parent element of $\beta$ is adjacent to the parent element of $\mathcal{A}$,
    \item[\textbf{Condition 2.}] $\mathcal{A}$ must intersect the physical domain of $\beta$ across the shared edge/face between the two elements (see Query 1 in Section~\ref{itm:query_1}),
    \item[\textbf{Condition 3.}] $\mathcal{A}$ must not be compatible with any enrichment in the set of ancestor enrichments of $\beta$, $\mathbb{A}_E^\beta$, and
    \item[\textbf{Condition 4.}] For each ancestor enrichment in $\mathbb{A}_E^{\mathcal{A}}$, if there is an enrichment in $\mathbb{A}_E^\beta$ that is compatible, then $\mathcal{A}$ and $\beta$ must both be on the positive side or both be on negative side of their respective ancestors.  There is one exception, if the two ancestor enrichments have opposite orientations, which can occur when cracks merge, then they should be on opposite sides of their respective ancestors.
\end{enumerate}

If $\mathcal{A}$ and $\beta$ meet all conditions, then it is topologically permissible for $\mathcal{A}$ to grow into $\beta$.  However, if the surface represents a crack, then there must also be a physics-based criterion to determine if the front should advance at enrichment $\mathcal{A}$. The criterion greatly depends on the type of crack model used.  For example, for the cohesive segment method, the criterion might dictate that the front should advance if the damage parameter in $\mathcal{A}$ is greater than a numerical tolerance.  As emphasized in the scope of this paper, we reserve detailed discussion of crack models for a future publication.

When the physics indicates that $\mathcal{A}$ should advance into $\beta$ and all four conditions above are met, we must determine those enrichments in adjacent elements with which the new enrichment, $\mathcal{B}$, should be compatible and determine the values of the signed distance function for $\mathcal{B}$ at the nodes of the parent element.  As previously discussed, the element-local level-set method does not require the signed distance function for two compatible enrichments to exactly match along the shared topology of the respective parent elements.  However, if enrichment $\mathcal{B}$ in element $e$ is to be compatible with some enrichment $\mathcal{C}$ in an adjacent element $n$, either
\begin{align}
sign\left(f^\mathcal{B}\left({\bar{x}}_{i_e}\right)\right) &= \phantom{+}sign\left(f^\mathcal{C}\left({\bar{x}}_{j_n}\right)\right)\;\forall\,{\bar{x}}_{i_e},{\bar{x}}_{j_n}\ |\ {\bar{x}}_{i_e}={\bar{x}}_{j_n}, \; \mathrm{or}
\label{eq:compat_req_a}
\\
sign\left(f^\mathcal{B}\left({\bar{x}}_{i_e}\right)\right) &= -sign\left(f^\mathcal{C}\left({\bar{x}}_{j_n}\right)\right)\;\forall\,{\bar{x}}_{i_e},{\bar{x}}_{j_n}\ |\ {\bar{x}}_{i_e}={\bar{x}}_{j_n}
\label{eq:compat_req_b}
\end{align}
must be satisfied, where ${\bar{x}}_{i_e}$ denotes the position of node $i$ in element $e$.  Therefore, the signed distance values affect whether $\mathcal{B}$ is permitted to be compatible with enrichments in adjacent elements, but the list of compatible enrichments also affects the signed distance function for $\mathcal{B}$, since the signed distance values must be determined such that either Eqn.~\ref{eq:compat_req_a} or Eqn.~\ref{eq:compat_req_b} is met.  To circumvent the circular dependency, we divide the algorithm into the following steps:

\begin{enumerate}[noitemsep,align=left, leftmargin=3.5em]
    \item[\textbf{Step 1.}] Construct a partial list of compatibilities for $\mathcal{B}$ that only includes the subset of enrichments that are compatible with any front enrichment of the surface to which $\mathcal{A}$ belongs and for which the parent element is adjacent to the parent element of $\mathcal{B}$, thus ensuring the new enrichment is compatible with the existing surface being advanced.
    \item[\textbf{Step 2.}] Find a temporary set of signed distance values for $\mathcal{B}$ that satisfy Eqn.~\ref{eq:compat_req_a} or Eqn.~\ref{eq:compat_req_b} based on the partial list of compatibilities from Step 1.
    \item[\textbf{Step 3.}] Use the temporary set of signed distance values from Step 2 to create the complete list of compatible enrichments for $\mathcal{B}$, checking if $\mathcal{B}$ should merge with any other enrichment in an adjacent element based on whether the difference of the signed distance values at the shared nodes is within a tolerance along the shared topology between the elements and any required physics-based criteria.
    \item[\textbf{Step 4.}] Determine the final signed distance values based on the complete list of compatible enrichments from Step 3.
\end{enumerate}

To determine either the temporary or final signed distance values, we follow the strategy proposed by Duan et al.\cite{Duan2009Element-local}, which we will describe here for completeness.  Given the set of enrichments that should be compatible with $\mathcal{B}$, $\mathbb{C}_E^\mathcal{B}$, let vector $\mathbf{b}$ contain a 1 or 0 for each node of the parent element of $\mathcal{B}$ (following the order of the element's connectivity) indicating whether the respective node is shared with the parent element of any enrichment in $\mathbb{C}_E^\mathcal{B}$.  Similarly, for each node of the parent element of $\mathcal{B}$, let $\mathbf{a}$ denote a vector containing the average signed distance value of all enrichments in $\mathbb{C}_E^\mathcal{B}$ that has a value at the respective node. For all nodes with a value of 1 in $\mathbf{b}$, the signed distance values will be weakly constrained to the corresponding value in $\mathbf{a}$. 
However, as Duan et al. recognized, $C^0$ continuity of the crack surface sometimes competes with the physics model governing the orientation of the crack surface.  Let $w_{C0}$ denote the relative weight of the $C^0$ continuity objective.  As $w_{C0}$ increases, the level-set for $\mathcal{B}$ tends towards $C^0$ continuity with the averaged level-sets for $\mathbb{C}_E^\mathcal{B}$.  Additionally, a system of equations at the element level can be solved to minimize the error of the weighted objective functions.  Let matrix $\mathbf{A}$ and vector $\mathbf{c}$ be given by
\begin{align}
A_{ij} &= \int{\frac{\partial\phi_i}{\partial x_k}\frac{\partial\phi_j}{\partial x_k}d\Omega_e}+\left(w_{C0}l_c\right)diag\left(\mathbf{b}\right)_{ij}\ \mathrm{and}
\label{eq:sd_solve_A}
\\
c_i &= \int{\frac{\partial\phi_i}{\partial x_j}{\hat{n}}_jd\Omega_e}+\left(w_{C0}l_c\right)a_i
\label{eq:sd_solve_c}
\end{align}
respectively, where $l_c$ is the cube root of the volume of the element and represents the characteristic length.  Then the signed distance values for $\mathcal{B}$ at each node, $\mathbf{f}^\mathcal{B}$, are given by $\mathbf{A}^{-1}\mathbf{c}$.  We start with $w_{C0}=1$ as recommended by the authors, but if Eqn.~\ref{eq:compat_req_a} and Eqn.~\ref{eq:compat_req_b} are violated, then $w_{C0}$ is increased by an order of magnitude and a new $\mathbf{f}^\mathcal{B}$ is determined.  This is repeated iteratively until either Eqn.~\ref{eq:compat_req_a} or Eqn.~\ref{eq:compat_req_b} is satisfied.  Since we also restrict the signed distance value at any node to be larger than a very small tolerance, this iterative procedure is guaranteed to converge.

\subsection{Numerical Integration}\label{integration}

In general, piecewise integration over the physical domain of each basal field, $\Omega_e^\beta$, and over the internal and external boundaries of each basal field, $\Gamma_{int}^\beta$ and $\Gamma_{ext}^\beta$, respectively, is needed to calculate the integrals appearing in Eqn.~\ref{eq:hhe_fea_equation}.  To numerically integrate over $\Omega_e^\beta$, we first apply Delaunay tetrahedralization to the polyhedron that approximates the physical domain of the field and then integrate over the tetrahedra.

\begin{figure}[ht!]
\centering
\includegraphics[height=1.6in]{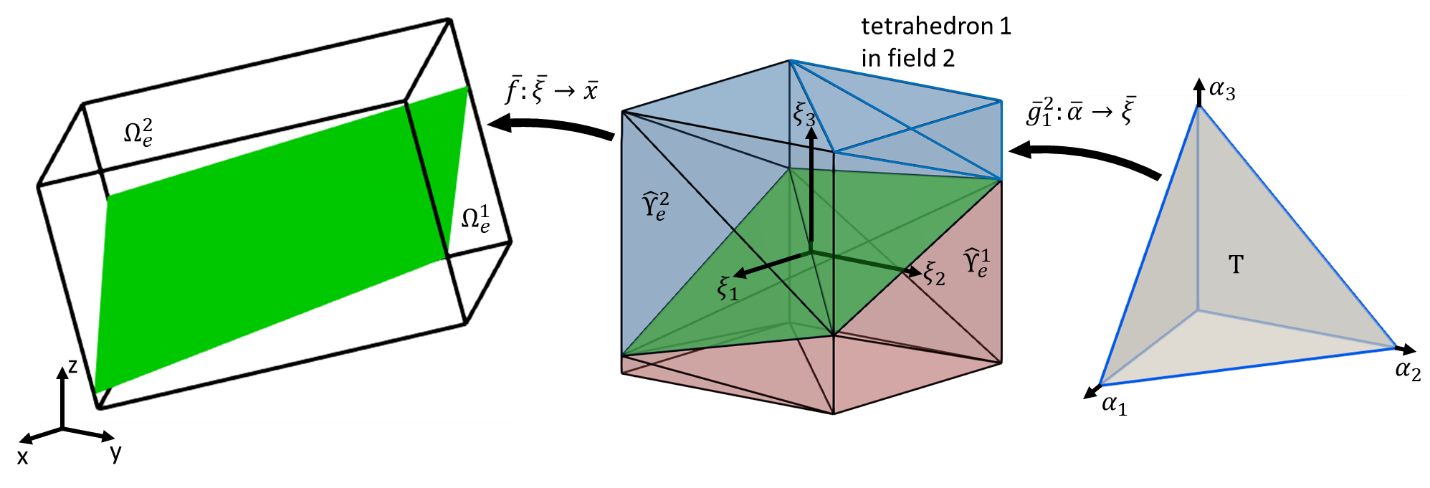}
\caption{Illustration of coordinate systems and maps between them.\label{fig:coord_maps}}
\end{figure}

There are three coordinate systems of interest.  Let $\bar{x}$ denote a coordinate in the reference frame of the global coordinate system, $\bar{\xi}$ denote the parametric coordinate system of the element, and $\bar{\alpha}$ denote the parametric coordinate system of a reference tetrahedron.  Additionally, let $\Omega_e^\beta$ denote the physical domain of field $\beta$ in $\bar{x}$, let $\Upsilon_e^\beta$ denote the physical domain in $\bar{\xi}$, let ${\hat{\Upsilon}}_e^\beta$ denote the approximation of $\Upsilon_e^\beta$ by a polyhedron, and let $T$ be the domain of a tetrahedron in $\bar{\alpha}$.  Figure~\ref{fig:coord_maps} illustrates the three coordinate systems and the maps between them.  Note that there is a $\bar{\alpha}\rightarrow\bar{\xi}$ map for each tetrahedron in the Delaunay tetrahedralization of each field, denoted by ${\bar{g}}_i^\beta$ for the $i^{th}$ tetrahedron of field $\beta$, although the figure only displays the 1st tetrahedron, outlined in blue. 

Integrating over ${\hat{\Upsilon}}_e^\beta$, the stiffness matrix appearing in Eqn.~\ref{eq:hhe_Kmatrix} becomes
\begin{align}
\mathbf{K}_\mathbf{e}^\beta = \int{\mathbf{B}^T\left(\bar{\xi}\right)\mathbf{C}\left(\bar{\xi}\right)\mathbf{B}\left(\bar{\xi}\right)\left|\frac{\partial\bar{f}}{\partial\bar{\xi}}\right|d{\hat{\Upsilon}}_e^\beta+\mathbf{\epsilon}^\beta},
\end{align}
where $\mathbf{\epsilon}$ is the discretization error arising due to either volumetric (measure) discretization errors and/or a mismatch between the integration space of our quadrature rule and the integrand.  The volumetric discretization error is nonzero in elements containing enrichments with level-sets that are not planar in $\bar{\xi}$.  Fortunately, elements with curved edges in the global coordinate system avoid additional discretization error, since the edges of the reference element are straight for the types of elements considered in this work.  Finally, we assume $\mathbf{\epsilon}^\beta$ is small and use the Delaunay tetrahedralization of ${\hat{\Upsilon}}_e^\beta$ to obtain
\begin{align}
\mathbf{K}_e^\beta\approx \sum_{i=1}^{M}{\int{\mathbf{B}^T\left(g_i^\beta\left(\bar{\alpha}\right)\right)\mathbf{C}\left(g_i^\beta\left(\bar{\alpha}\right)\right)\mathbf{B}\left(g_i^\beta\left(\bar{\alpha}\right)\right) \left|\frac{\partial{\bar{g}}_i^\beta\left(\bar{\alpha}\right)}{\partial\bar{\alpha}}\right|\left|\frac{\partial\bar{f}\left(\bar{\xi}\right)}{\partial\bar{\xi}}\right| dT}}, \label{eq:hhe_Kmatrix_integrated}
\end{align}
where M denotes the number of tetrahedra for the field.  

Depending on the type of element, we select a Gaussian quadrature scheme for the tetrahedra to exactly integrate the integrand when $\mathbf{C}$ is constant throughout the element.  For example, for a linear hexahedron element, the integrand is quadratic with respect to $\bar{\xi}$, so a 5-point Gaussian quadrature scheme specialized for the tetrahedra exactly integrates the integrand in Eqn.~\ref{eq:hhe_Kmatrix_integrated}. We can use a similar strategy for integrating over $\Gamma_{int}^\beta$ and $\Gamma_{ext}^\beta$.  However, since the verification cases shown in this paper only consider traction-free enrichment surfaces, we will reserve further discussion on integration strategy for these terms for future work.

\subsection{Degree of Freedom Enumeration via Basal Field Compatibilities}\label{dof_enumeration}

For the continuous-Galerkin (CG) method, the displacement field should remain continuous everywhere, except across enrichment surfaces that model a discontinuity.  To enforce continuity of basal fields across elements, we enumerate the degrees of freedom associated with a basal field in one element with the same numbers at the shared nodes of the adjacent element.  Enumerating the DoFs this way ensures that all basal fields are continuous across elements without the need for constraints in the system of equations.  However, because we allow the implicit and explicit representations of enrichment surfaces to be discontinuous across elements, we cannot always enforce continuity of aggregate fields.  In other words, the aggregated solution may be discontinuous across a face if the enrichment surface is discontinuous across the two connected elements.  We could ameliorate this by enforcing continuity of the enrichment surfaces at the cost of artificially constraining how enrichment surfaces can grow.  However, the discontinuity of the enrichment surface across adjacent elements is typically very small.

When a field is enriched, an additional set of DoFs must be introduced for the element, resulting in two new basal fields.  Rather than update the DoF enumeration every time an element is enriched, we elect to track new enrichments that occur and renumber the DoFs for all elements when an updated solution is needed.  However, generating a suitable DoF enumeration directly from the enrichment trees within all elements is an expensive operation, since determining the DoFs for a basal field in an element relies on complex logic involving the enrichment trees of the given element and all of its neighbors.  Instead, we propose storing a graph of compatibilities between basal fields in adjacent elements, which can be incrementally updated to accommodate new enrichments.  The existence of this compatibility graph allows for an efficient DoF enumeration of the entire domain when an updated solution is required.

In this section, we describe the algorithm for constructing and updating the basal field compatibility graph and the DoF enumeration algorithm that naturally maintains a CG solution across elements, including how the algorithm is suitable for distributed computing.

\subsubsection{Basal Field Compatibility Graph}

As discussed in a previous section, a \emph{compatibility} between a basal field in one element and a basal field in the adjacent element means that the solution field should have $C^0$ continuity within those basal fields across the two elements.  For the simplest case, consider an enrichment surface that cuts through two elements.  The two basal fields on the positive side of the surface should be compatible with each other, and separately, the two basal fields on the negative side should also be compatible with each other.  Whether two basal fields should be compatible can be deduced from the enrichment trees in the two elements, topological information, and the pairs of compatible enrichments between the two elements.  

\begin{figure}[ht!]
    \centering
    \begin{subfigure}[t]{0.45\textwidth}
        \centering
        \includegraphics[height=1.6in]{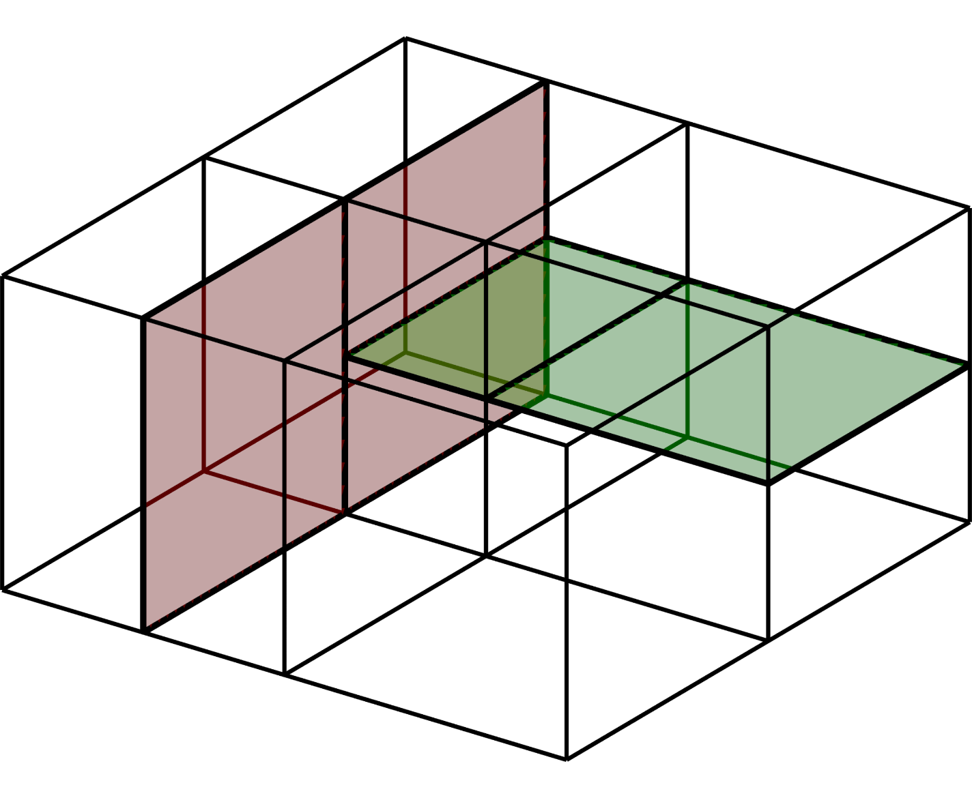}
        \caption{Orthogonal enrichment surfaces illustrating compatibility cases.}\label{fig:compatibility_cases_a}
    \end{subfigure} %
    ~ 
    \begin{subfigure}[t]{0.45\textwidth}
        \centering
        \includegraphics[height=1.6in]{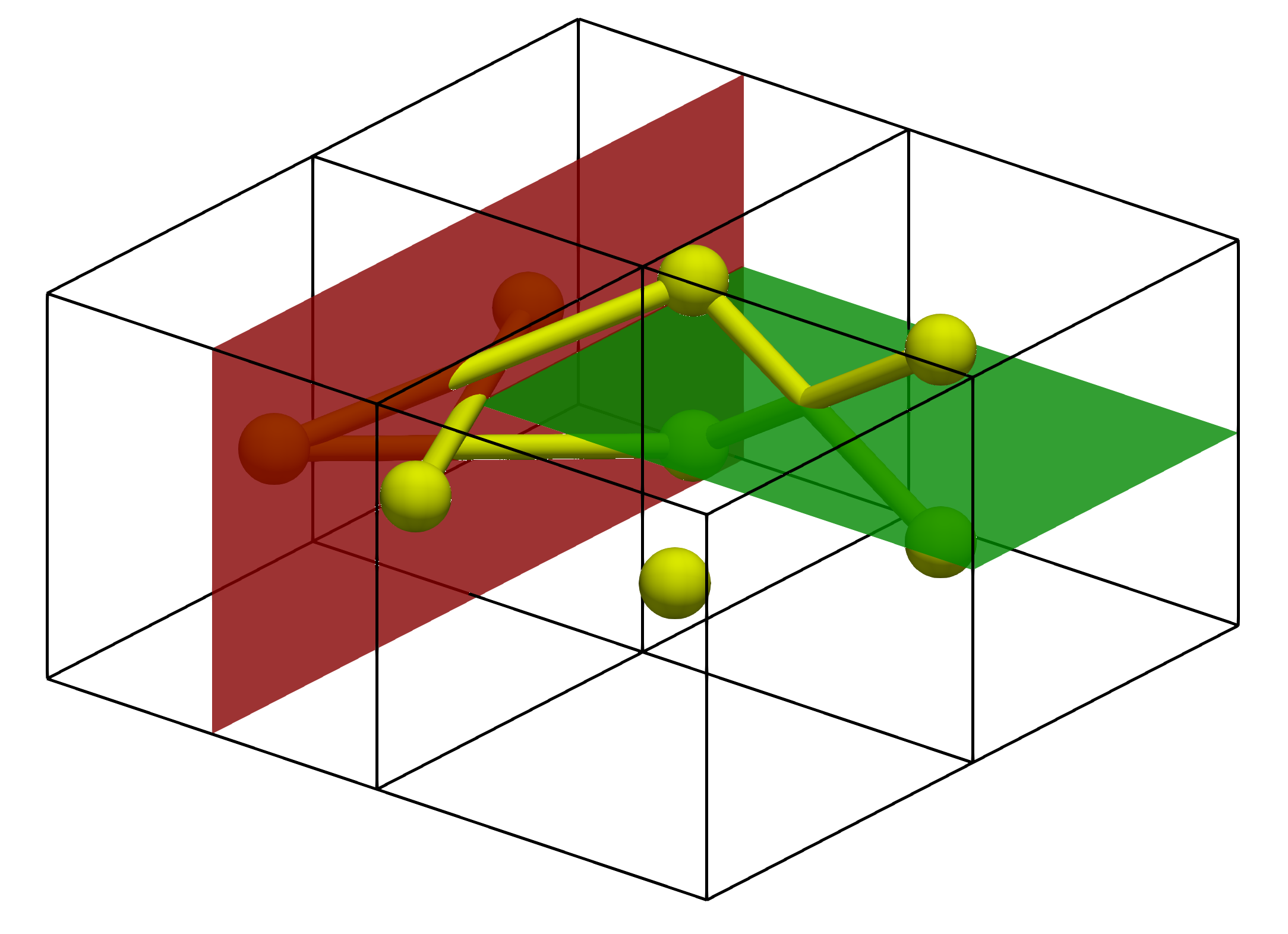}
        \caption{Some of the incompatibilities between basal fields across adjacent elements.}\label{fig:compatibility_cases_b}
    \end{subfigure} %
    ~ 
    \begin{subfigure}[t]{0.45\textwidth}
        \centering
        \includegraphics[height=1.6in]{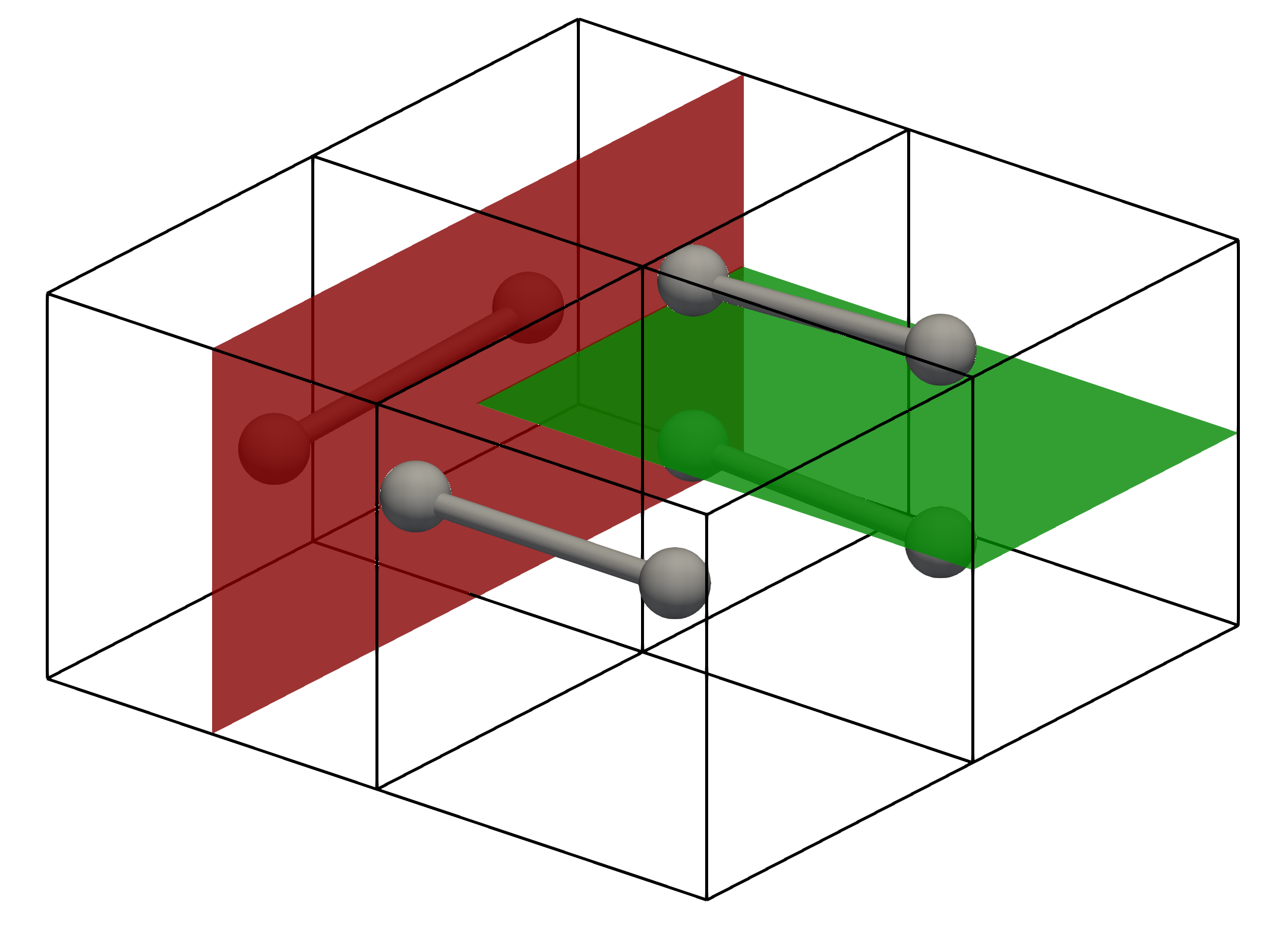}
        \caption{One-to-one compatibilities between basal fields across adjacent elements.}\label{fig:compatibility_cases_c}
    \end{subfigure} %
    ~ 
    \begin{subfigure}[t]{0.45\textwidth}
        \centering
        \includegraphics[height=1.6in]{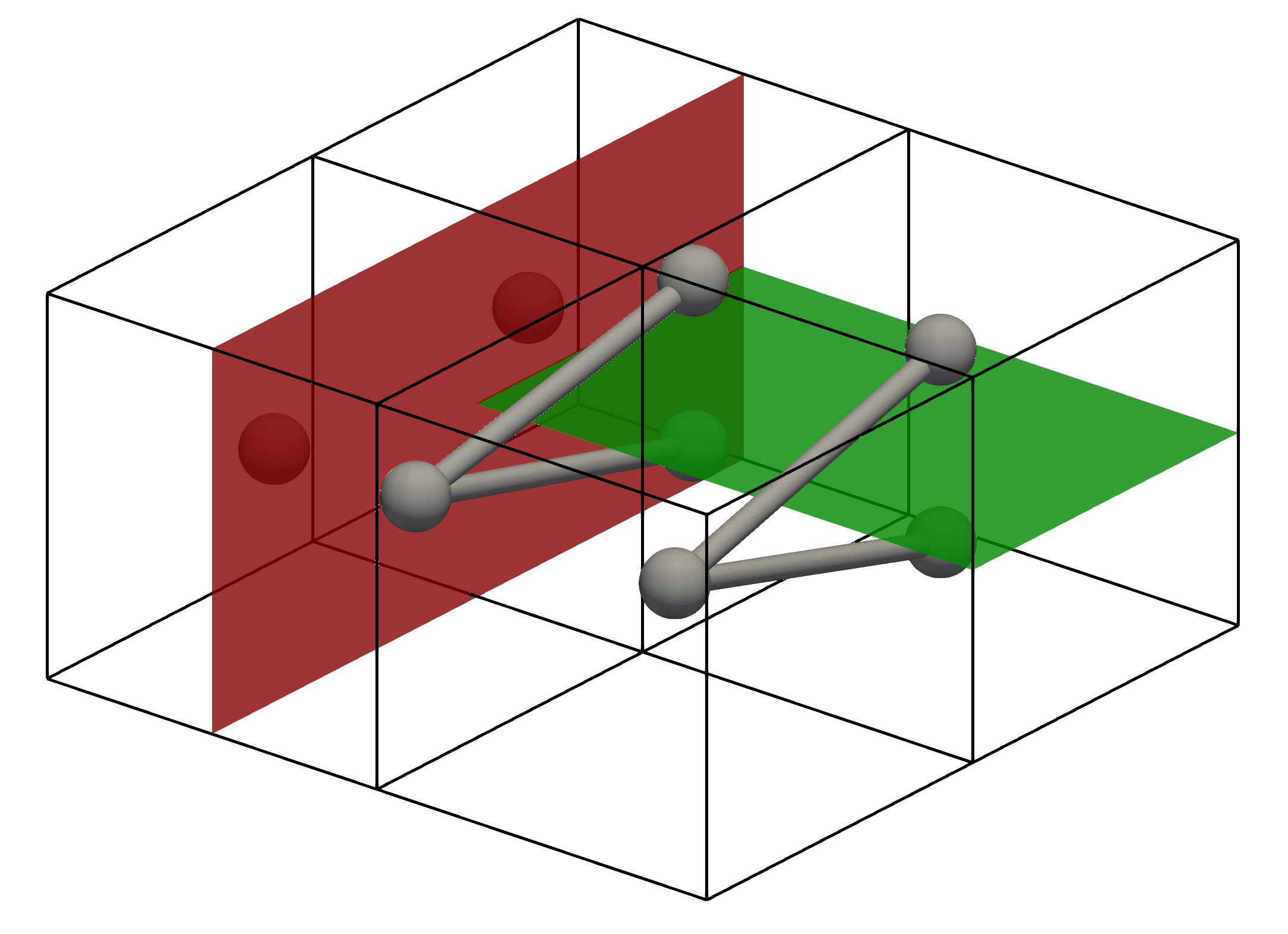}
        \caption{Many-to-one compatibilities between basal fields across adjacent elements.}\label{fig:compatibility_cases_d}
    \end{subfigure}
    \caption{Illustration of all three cases pertaining to compatibilities of basal fields.\label{fig:compatibility_cases}}
\end{figure}

First, it is helpful to delineate three types of relationships between a basal field in one element and the basal fields in an adjacent element.  The given basal field can either be: 1) incompatible with all basal fields in the adjacent element, 2) compatible with exactly one basal field in the adjacent element, or 3) compatible with multiple basal fields in the adjacent element.  Fig.~\ref{fig:compatibility_cases} illustrates a situation with all three types.  In Fig.~\ref{fig:compatibility_cases_b}-\ref{fig:compatibility_cases_d}, we represent each basal field with a sphere, which is placed at the centroid of the physical domain of the basal field.  Additionally, we indicate a compatibility between two basal fields with a grey line connecting the two spheres, and similarly, we represent an incompatibility between two basal fields with a yellow line.  The DoF enumeration will enforce continuity of the solution across any compatible basal fields in adjacent elements for which the compatibility appears in the graph.  Consequently, there is a choice of which compatibilities to include.  Some crack models weakly enforce continuity of the solution along the front, including the cohesive segment method since it requires a closed cohesive element along the front.  In this case, only including compatibilities involving exactly one basal field in one element and exactly one basal field in the adjacent element, such as those shown in Fig.~\ref{fig:compatibility_cases_c}, might lead to a more well-conditioned system of equations.  However, other crack models might require the DoF enumeration to create continuity of the solution along the front, in which case all compatibilities should be included, but as a reminder, something must be done to address the stress singularity if the crack model does not smear or regularize the effect of the crack tip.  For the verification cases shown later in this paper, the choice is irrelevant since we will restrict the verification to situations where the front only terminates at another enrichment surface or the boundary of the model, which precludes the possibility of compatibilities involving one basal field in one element and multiple basal fields in the adjacent element, such as those shown in Fig.~\ref{fig:compatibility_cases_d}.

\clearpage

\begin{figure}[ht!]
    \centering
    \begin{subfigure}[t]{0.45\textwidth}
        \centering
        \includegraphics[width=0.8\textwidth]{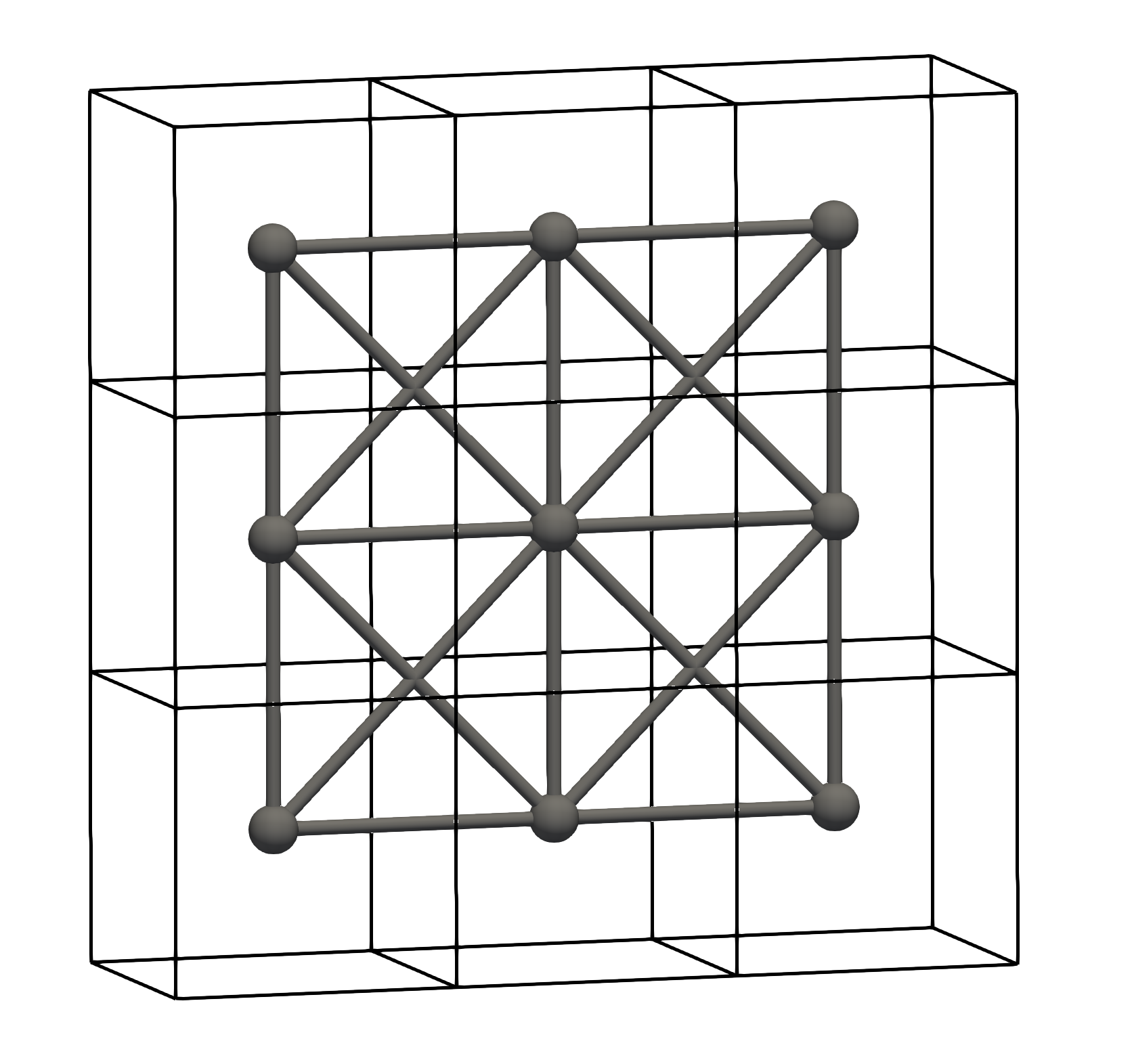}
        \caption{Initial compatibility graph for 3x3x1 uniform grid.}\label{fig:compatibility_map_a}
    \end{subfigure} %
    ~ 
    \begin{subfigure}[t]{0.45\textwidth}
        \centering
        \includegraphics[width=0.8\textwidth]{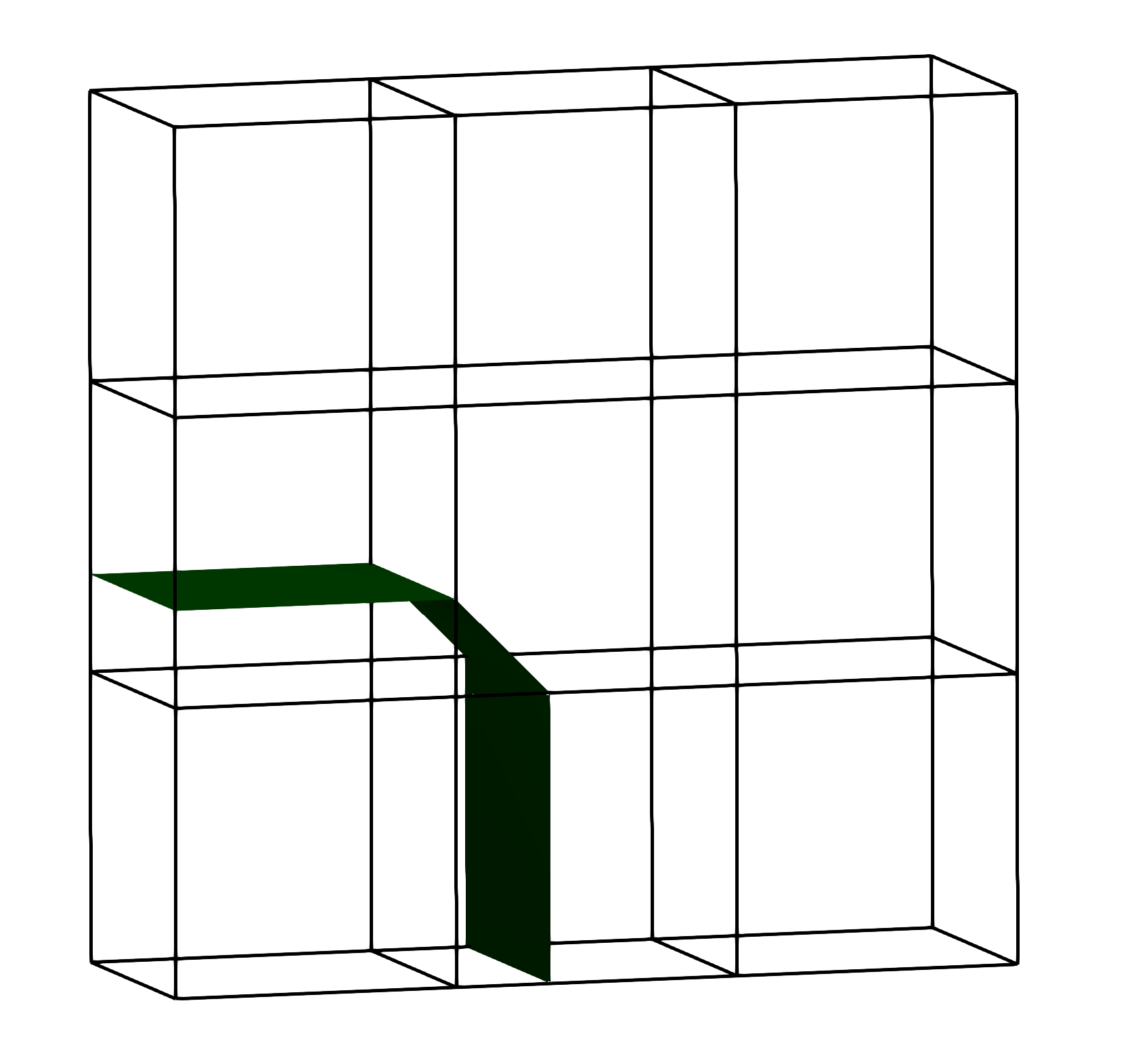}
        \caption{Hypothetical enrichment surface to insert.}\label{fig:compatibility_map_b}
    \end{subfigure} %
    ~
    \begin{subfigure}[t]{0.45\textwidth}
        \centering
        \includegraphics[width=0.8\textwidth]{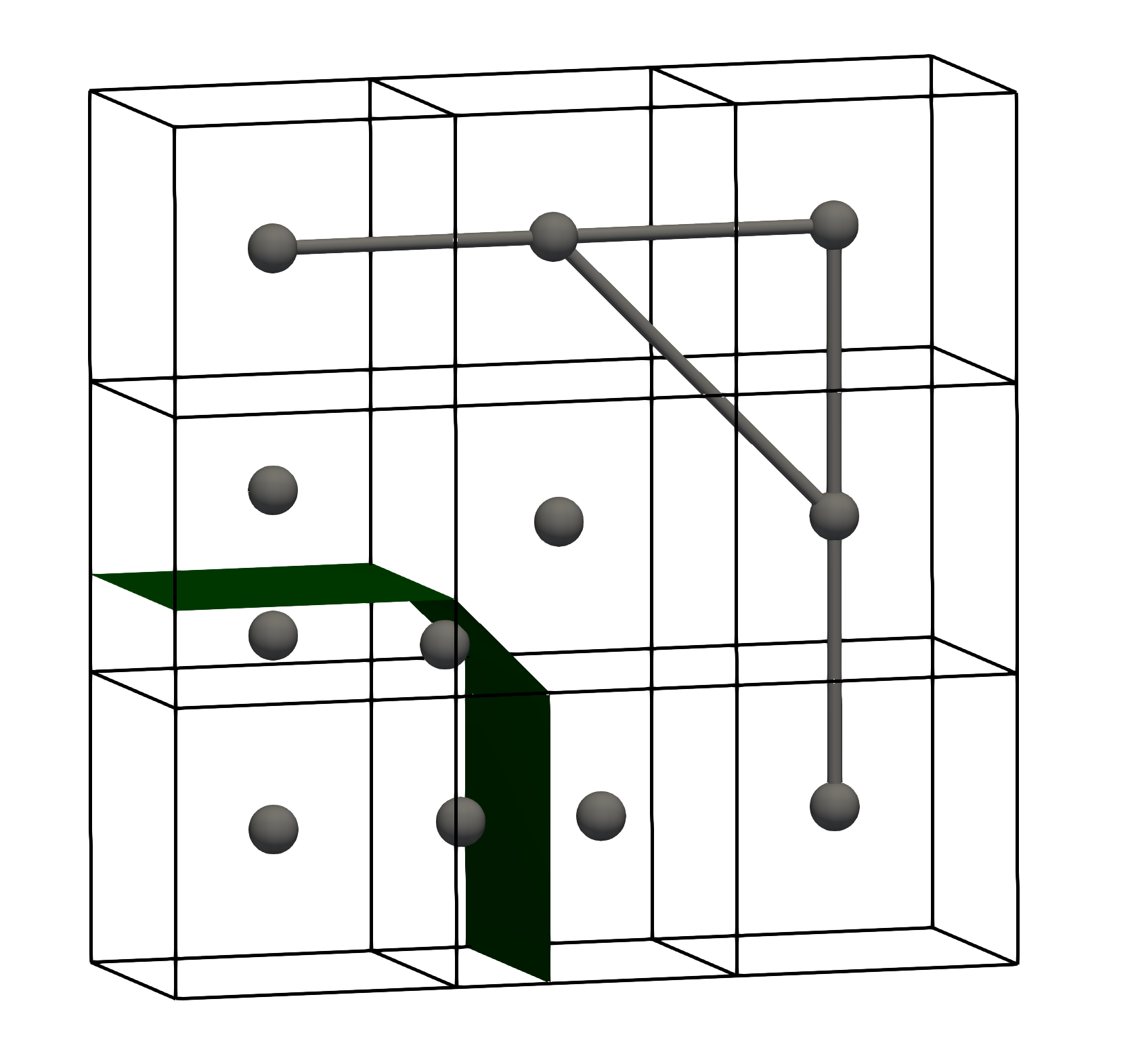}
        \caption{After removal of compatibilities involving recently enriched basal fields.}\label{fig:compatibility_map_c}
    \end{subfigure} %
    ~
    \begin{subfigure}[t]{0.45\textwidth}
        \centering
        \includegraphics[width=0.8\textwidth]{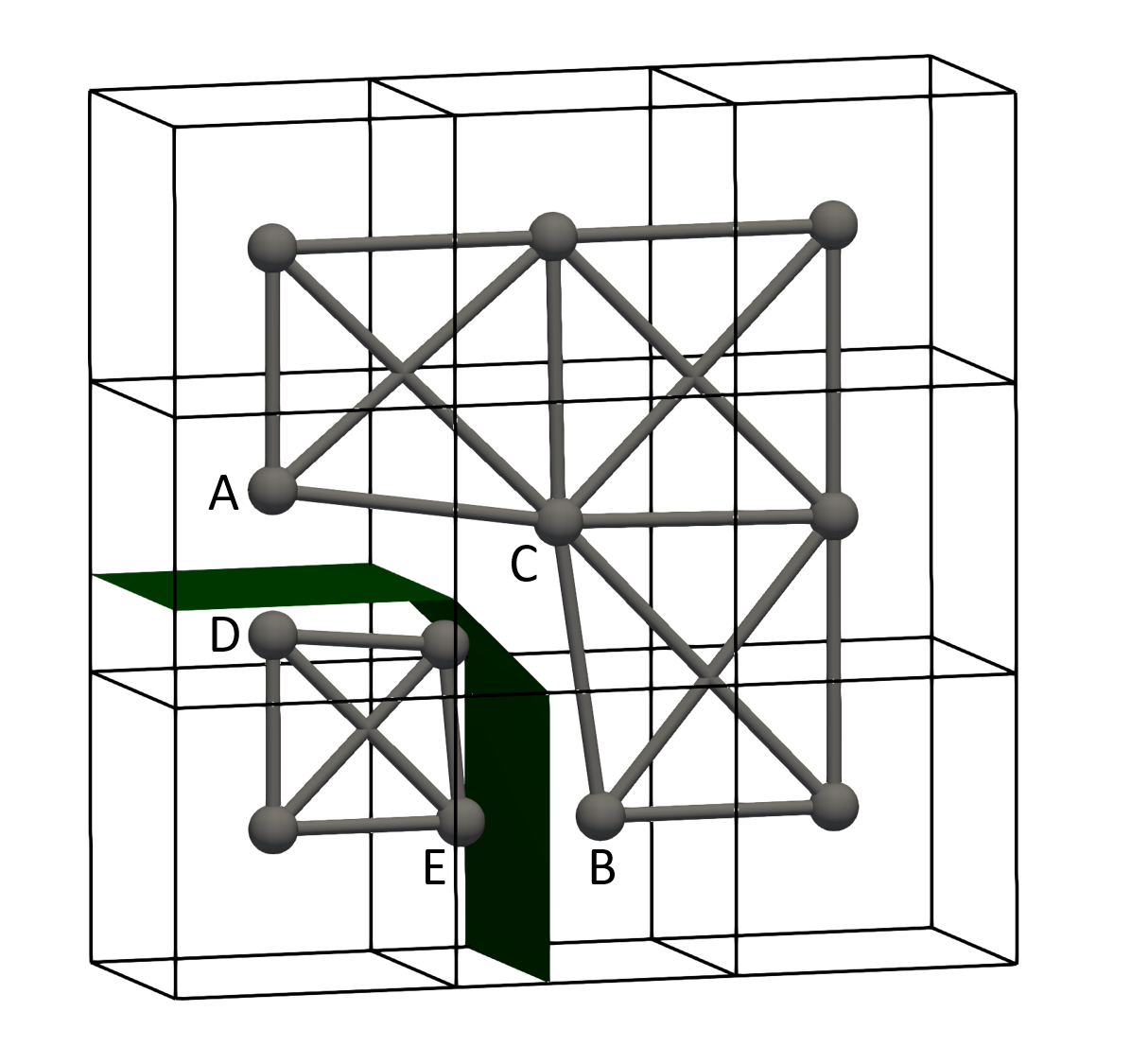}
        \caption{Partial compatibility graph based on enrichment trees and element topology.}\label{fig:compatibility_map_d}
    \end{subfigure} %
    ~
    \begin{subfigure}[t]{0.45\textwidth}
        \centering
        \includegraphics[width=0.8\textwidth]{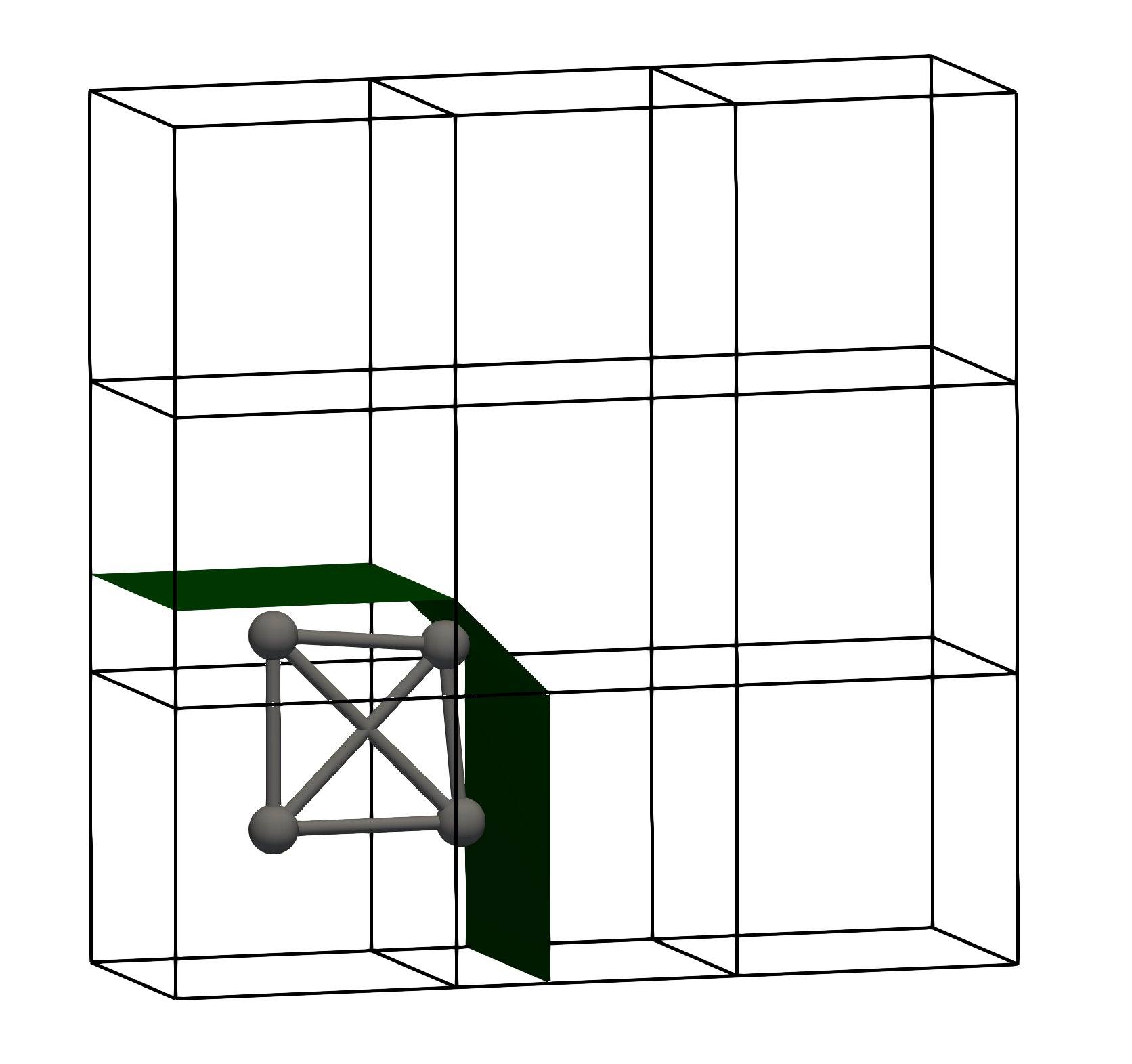}
        \caption{Updated compatibility graph (part 1).}\label{fig:compatibility_map_e}
    \end{subfigure} %
    ~
    \begin{subfigure}[t]{0.45\textwidth}
        \centering
        \includegraphics[width=0.8\textwidth]{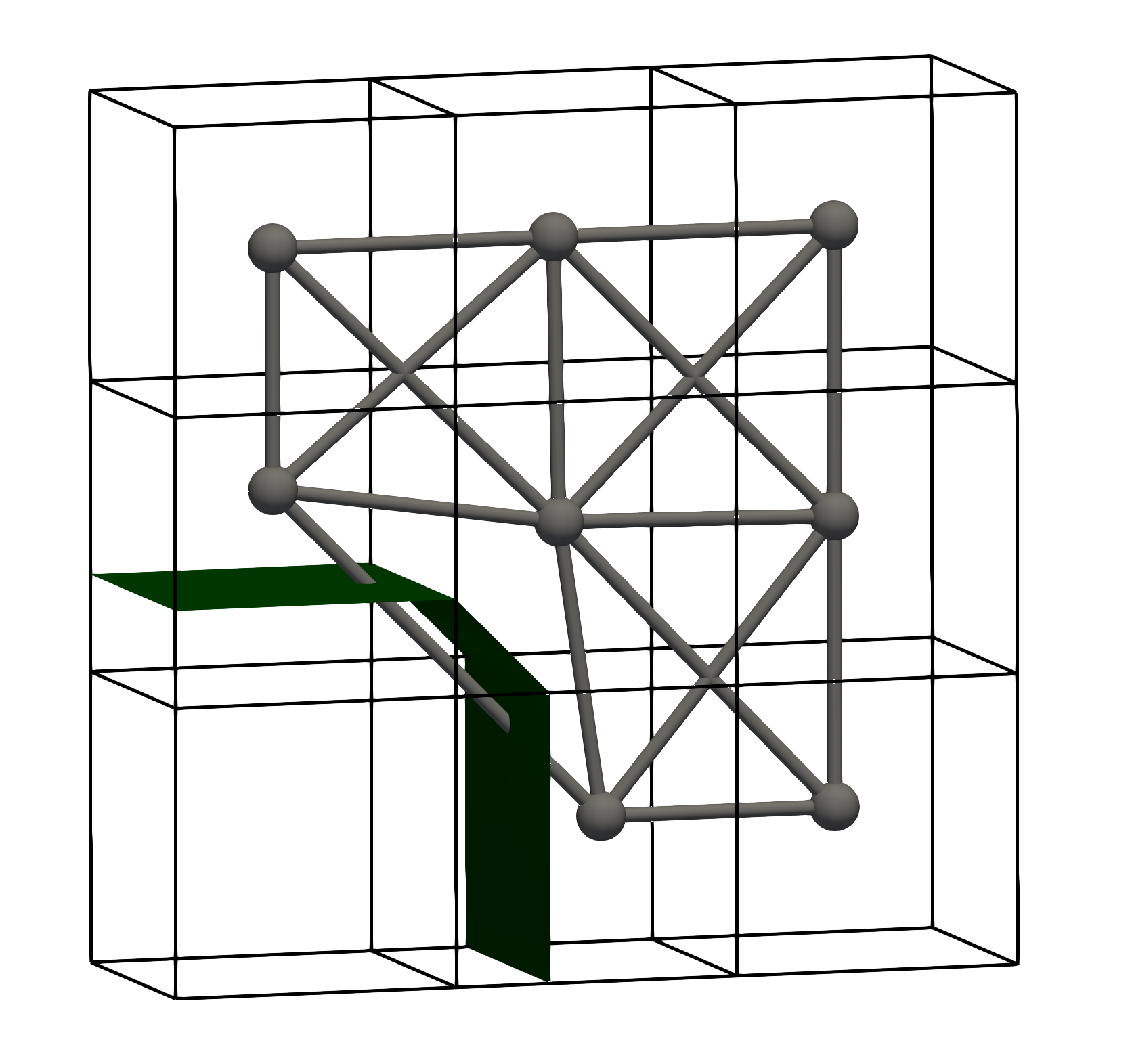}
        \caption{Updated compatibility graph (part 2).}\label{fig:compatibility_map_f}
    \end{subfigure}
    \caption{Illustration of initial compatibility graph map and the steps update the graph to account for new enrichments.\label{fig:compatibility_map}}
\end{figure}

Storing compatibilities between basal fields in adjacent elements in a graph is convenient for a computer implementation, and before any enrichments are introduced, the compatibility graph is equivalent to the element adjacency list.  For example, Fig.~\ref{fig:compatibility_map_a} shows a 3x3x1 grid and the graph of compatibilities between basal fields.  When a new enrichment is introduced in an element, the compatibilities between basal fields in the neighborhood of the enriched element are disrupted and must be updated.  However, all compatibilities involving elements not adjacent to any enriched element remain unaffected.  This allows the compatibility graph to be updated only in the neighborhood of new enrichments when an updated DoF enumeration is required.  For example, when the enrichment surface shown in Fig.~\ref{fig:compatibility_map_b} is inserted, then compatibilities involving the enriched elements are removed from the graph, as illustrated in Fig.~\ref{fig:compatibility_map_c}. After all enrichments have been inserted and a new DoF enumeration is required, the correct compatibilities are added back using an algorithm that requires two passes.  In the first pass, a partial compatibility graph is constructed based on the element topology and enrichment tree information.  This partial compatibility graph may exclude some corner cases.  Consequently, in the second pass, the element topology and partial compatibility graph are used to find a complete compatibility graph.

To create the partial compatibility graph, a compatibility should be added between basal fields $\beta$ and $\gamma$ if:
\begin{enumerate}[noitemsep,align=left, leftmargin=3.5em]
    \item[\textbf{1.}] The parent element of $\beta$ is adjacent to the parent element of $\gamma$,
    \item[\textbf{2.}] Both $\beta$ and $\gamma$ touch the shared edge (if the two elements share an edge) or the shared face (if the two element share a face), and
    \item[\textbf{3.}] For each ancestor enrichment in $\mathbb{A}_E^\beta$, if there is an enrichment in $\mathbb{A}_E^\gamma$ that is compatible, then $\beta$ and $\gamma$ must both be on the positive side or both be on negative side of their respective ancestors.  There is one exception, if the two ancestor enrichments have opposite orientations, which can occur when cracks merge, then they should be on opposite sides of their respective ancestors.
\end{enumerate}

Note that the above conditions describe the relationship between two basal fields.  While similar, these conditions differ from the enrichment growth conditions in Section~\ref{growth_algorithm}, which relate an enrichment to an adjacent basal field.

As previously discussed, one might elect to exclude compatibilities for a basal field involving multiple basal fields in the adjacent element, and whether a basal field is compatible with a single basal field or multiple basal fields in the adjacent element can be trivially deduced from the compatibility graph after the fact.  However, based on the information available before the compatibility graph is updated, $\beta$ can be guaranteed to only be compatible with $\gamma$ out of the set of basal fields in the parent element of $\gamma$ and vice versa if all of the following are met in addition to the three conditions above:
\begin{enumerate}[noitemsep,align=left, leftmargin=3.5em]
    \item[\textbf{1.}] No enrichment in $\mathbb{A}_E^\beta$ is compatible with multiple enrichments in $\mathbb{A}_E^\gamma$,
    \item[\textbf{2.}] No enrichment in $\mathbb{A}_E^\gamma$ is compatible with multiple enrichments in $\mathbb{A}_E^\beta$,
    \item[\textbf{3.}] Any enrichment in $\mathbb{A}_E^\beta$ that intersects the shared topology between the adjacent elements is compatible with an enrichment in $\mathbb{A}_E^\gamma$, and
    \item[\textbf{4.}] Any enrichment in $\mathbb{A}_E^\gamma$ that intersects the shared topology between the adjacent elements is compatible with an enrichment in $\mathbb{A}_E^\beta$.
\end{enumerate}

By testing these conditions, we can construct a partial compatibility graph, such as the one shown in Fig.~\ref{fig:compatibility_map_d}.  However, as mentioned, this partial compatibility graph excludes a corner case, namely, a compatibility between the labelled fields $A$ and $B$.  The compatibility is not included automatically because the element topology and enrichment tree information are insufficient.  Note that the compatibility between $D$ and $E$ is found because the two fields touch the shared topology between the respective parent elements.  However, $A$ and $B$ are compatible through a basal field, $C$, that lies in an element adjacent to the parents of both $A$ and $B$, and this type of compatibility cannot be deduced from the enrichment trees of the two elements alone.  After the partial compatibility graph is constructed, a second pass over the elements and basal fields is performed to ensure that a compatibility exists between any pair of basal fields, $\beta$ and $\gamma$, that lie in adjacent elements for which there is an $\alpha$ that is compatible with both $\beta$ and $\gamma$.  For example, Fig.~\ref{fig:compatibility_map} shows the complete compatibility graph after the second pass.  As it is difficult to visually distinguish, we present this graph in two parts, one for each side of the enrichment surface (Fig.~\ref{fig:compatibility_map_e} and \ref{fig:compatibility_map_f}).

\clearpage

Importantly, this algorithm for creating the basal field compatibility graph is embarrassingly parallel.  In a multithreaded paradigm, each thread can be assigned a subset of elements and fill in the compatibility graph for the assigned elements, although adjacent elements assigned to different threads will be visited multiple times.  Storing the graph within an adjacency list ensures thread-safety if each thread only modifies the list of compatibilities for the assigned elements.  

Within a distributed paradigm combined with domain decomposition, the mesh is partitioned, with each element and each node assigned to a particular rank.  Each rank may only have access to the subset of the mesh that contains all owned elements and any element containing an owned node.  Let a ghost element or node refer to an element or node that is accessible in a partition of the mesh but is not owned.   As long as each rank has the topological and enrichment tree information for each ghost element in its partition, then no communication between ranks is required for each rank to build a compatibility graph for its partition.

\clearpage

\subsubsection{Degree of Freedom Enumeration}

In our proposed XFEM approach, each degree of freedom (DoF) corresponds to a component of the solution vector at a particular nodal position and receives a number indicating the corresponding row in the global system of equations.  Consequently, it is convenient to store a map that takes an element, basal field, and nodal position and returns the DoF numbers.  In turn, the DoF map can be used to efficiently assemble element level matrices/vectors into global matrices/vectors.  In our implementation, the DoF map is stored as an array of DoF numbers ordered first by element, then by basal field, and then by nodal position relative to the respective element's connectivity.  To maintain a CG solution, the DoF enumeration must ensure that every basal field has the same DoF number at a nodal position as another compatible basal field at the same nodal position.  Of course, there are many enumerations that would satisfy this requirement.  We propose an algorithm for cheaply creating an admissible DoF enumeration given the basal field compatibility graph.  The resulting enumeration can then be combined with one of the many established methods for permuting the enumeration to improve solver performance, such as METIS\cite{Metis1998}.  

Our approach for cheaply creating an admissible DoF enumeration is to loop over the elements and the basal fields within the element, assigning a DoF index to each nodal position following
\begin{align}
    d[e, \beta, i]=
    \begin{cases}
        d[n, \gamma, j], &\mathrm{if\ } d[n, \gamma, j] \mathrm{\ is\ initialized\ } | \ \gamma \in \mathbb{C}_E^\mathcal{B} \mathrm{\ and\ } {\bar{x}}_{i_e}={\bar{x}}_{j_n}, \\
        \mathrm{new\ index}, &\mathrm{otherwise,}
    \end{cases}
    \label{eq:dof_assignment}
\end{align}
where $d[e, \beta, i]$ denotes the DoF index for basal field $\beta$ in element $e$ at nodal position $i$ and $\mathbb{C}_F^\mathcal{\beta}$ denotes the set of basal fields compatible with $\beta$ in any adjacent element $n$.

Similar to the compatibility graph, this enumeration algorithm is easily parallelized through domain decomposition.  If the mesh is partitioned with each element and each node assigned to a particular rank.  A local, admissible DoF enumeration within a rank can be created in the following two steps:
\begin{enumerate}
    \item Loop over the elements (both ghost and owned) and the basal fields within each element, assigning DoF indices per Eqn.~\ref{eq:dof_assignment} for owned and ghost nodal positions separately with each range starting at 0, and then
    \item Revisit positions of the DoF map corresponding to a ghost nodal position and increment the index by the largest index at any owned nodal position.
\end{enumerate}

Importantly, the indices for DoFs corresponding to the owned nodal positions will appear at the front of the local enumeration, followed by those corresponding to ghost nodal positions.  Next, a local-to-global map can be constructed by each rank as follows:
\begin{enumerate}
    \item Perform a parallel prefix sum (scan) of the number of DoFs assigned to owned nodal positions, which each rank can use to determine their starting global index,
    \item Populate the first section of the local-to-global map that corresponds to the owned nodal positions with increasing integers starting at the starting global index for the rank,
    \item Send the global indices for any owned nodal position that appears as a ghost node on another rank to that rank, and
    \item Complete the local-to-global map using the list of global indices for all ghost nodal positions that were received in the previous step.
\end{enumerate}

The only communication required to create a compatibility graph and DoF enumeration occurs in steps 1 and 3.  Step 1 is a collective operation that is known to run in $O(log\ n)$ time, while step 3 involves communication between each rank and the other ranks whose partition borders its own partition.  A deeper discussion of the details of a parallel implementation and a characterization of its scalability will be reserved for a future article.

\section{Verification}\label{verification}

Thus far, we described the algorithms and implementation of our CG-XFEM approach.  In this section, we verify our complete FEA implementation via a few linear elastic analyses, focusing on certain challenging cases.  For the remainder of this section, we will model open cracks using enrichment surfaces.   In many applications, a cohesive crack model would allow crack fronts to exist within the domain without introducing stress singularities.  However, since cohesive crack models and surface integration techniques are outside the scope of this paper, we only consider verification cases where enrichment surfaces evolve until the fronts of the surfaces encounter a mesh boundary, encounter another enrichment surface, or create a closed manifold. As a consequence of this choice, enrichment surfaces divide the computational domain into multiple disconnected pieces.  We visualize the solutions using a custom ParaView\cite{ParaView} plugin tailored for XFEM data.  The verification cases were selected to highlight two features of our method.  First, they show that our DoF enumeration algorithm correctly allows a discontinuity in the displacement field along enrichment surfaces and maintains a continuous displacement field elsewhere.  Second, they illustrate that our method is not restricted to uniform grids and hexahedral elements.

\subsection{Intersection of Curved Enrichment Surfaces Within a Mixed-Element Mesh}\label{verification_1}

\begin{figure}[ht!]
\centering
    \begin{subfigure}[t]{0.4\textwidth}
        \centering
        \includegraphics[height=1.8in]{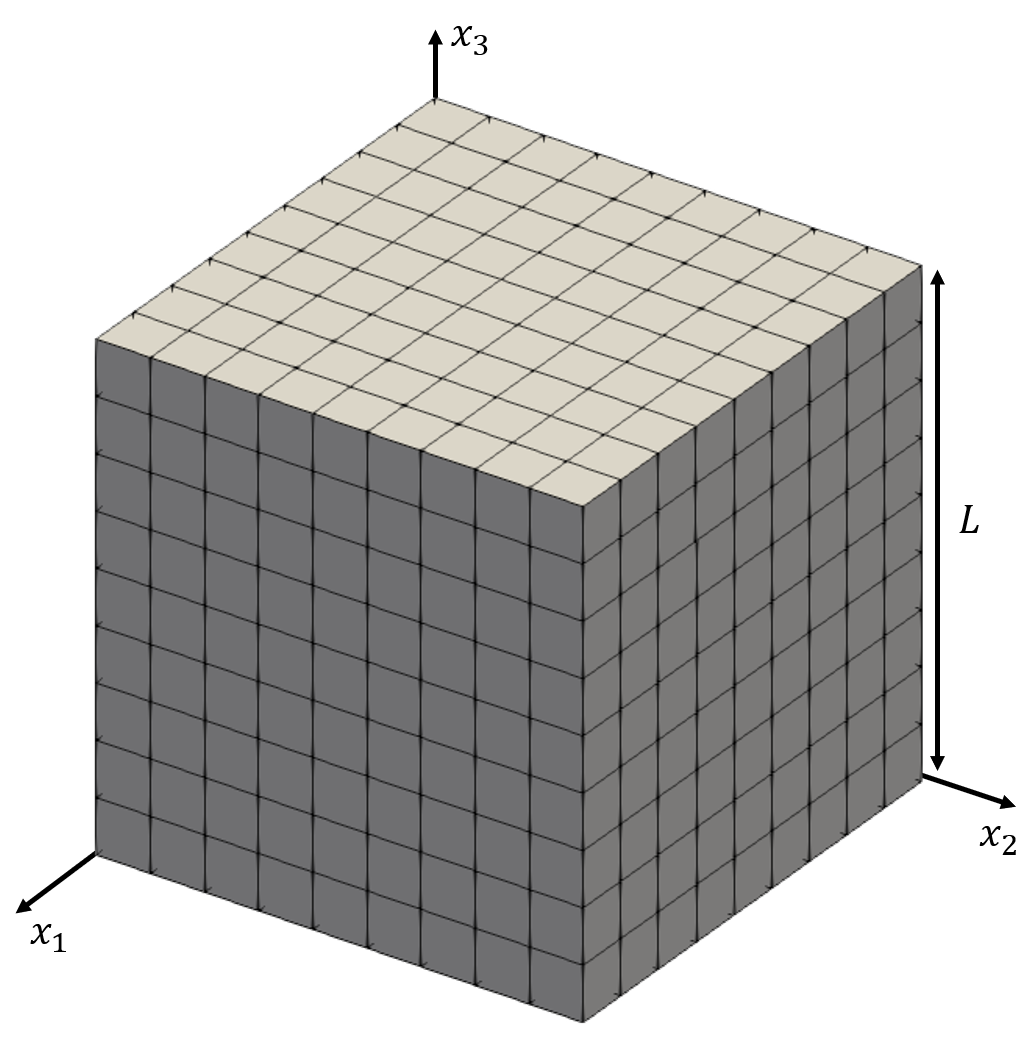}
        \caption{Mixed-element mesh with axes, side length labeled, and element edges shown\label{fig:ver_curved_a}}
    \end{subfigure} %
    ~ 
    \begin{subfigure}[t]{0.4\textwidth}
        \centering
        \includegraphics[height=1.8in]{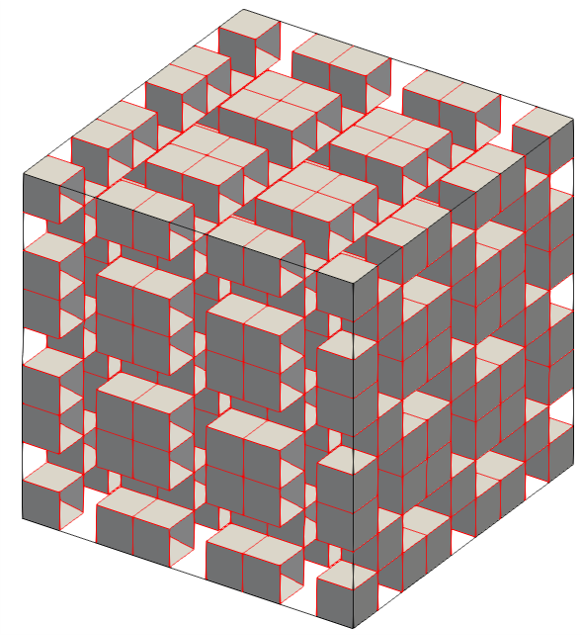}
        \caption{The 1350 pyramid elements within the mixed-element mesh}
    \end{subfigure}

    \begin{subfigure}[t]{0.4\textwidth}
        \centering
        \includegraphics[height=1.8in]{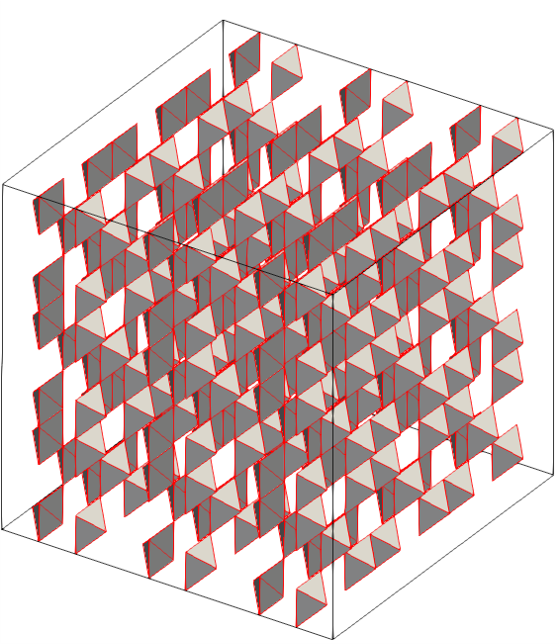}
        \caption{The 540 tetrahedron elements within the mixed-element mesh}
    \end{subfigure} %
    ~ 
    \begin{subfigure}[t]{0.4\textwidth}
        \centering
        \includegraphics[height=1.8in]{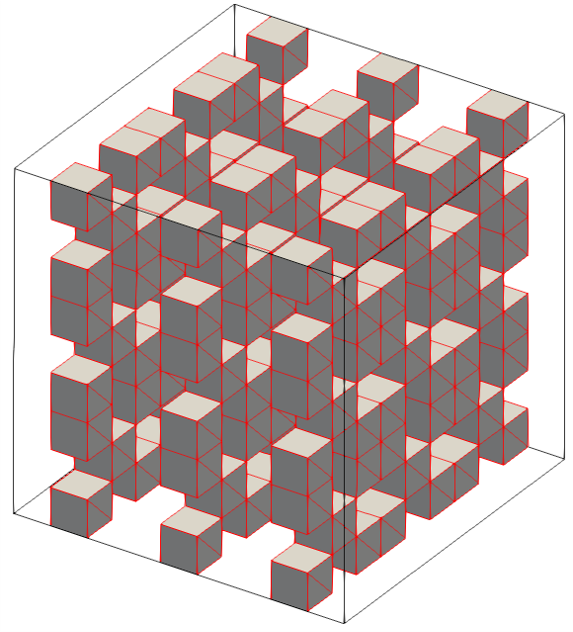}
        \caption{The 270 wedge elements within the mixed-element mesh}
    \end{subfigure}

    \begin{subfigure}[t]{0.4\textwidth}
        \centering
        \includegraphics[height=1.8in]{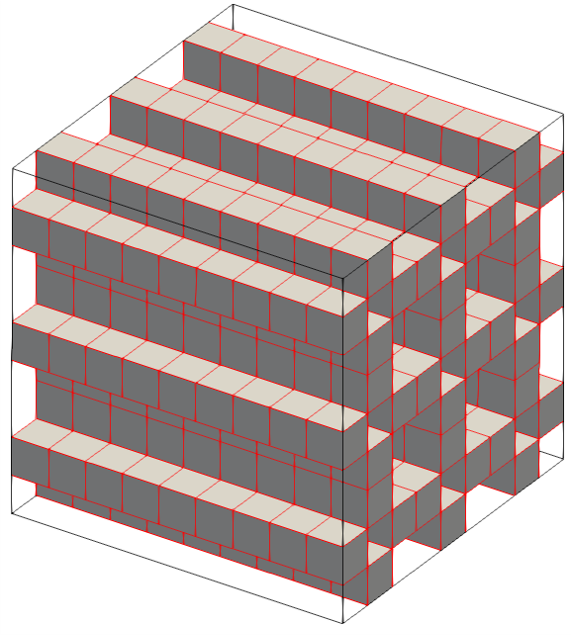}
        \caption{The 320 hexahedron elements within the mixed-element mesh}
    \end{subfigure} %
    \caption{Mixed-element mesh used for the verification case of intersecting curved enrichment surfaces.\label{fig:ver_curved_mesh}}
\end{figure}

For this verification case, we consider two intersecting curved enrichment surfaces within a mixed-element mesh, consisting of a combination of pyramid, wedge, hexahedron, and tetrahedron elements.  Figure~\ref{fig:ver_curved_mesh} shows the mesh, which has a side length $L$, and a view of the elements belonging to each type.

\begin{figure}[ht!]
    \centering
    \begin{subfigure}[t]{0.4\textwidth}
        \centering
        \includegraphics[height=1.8in]{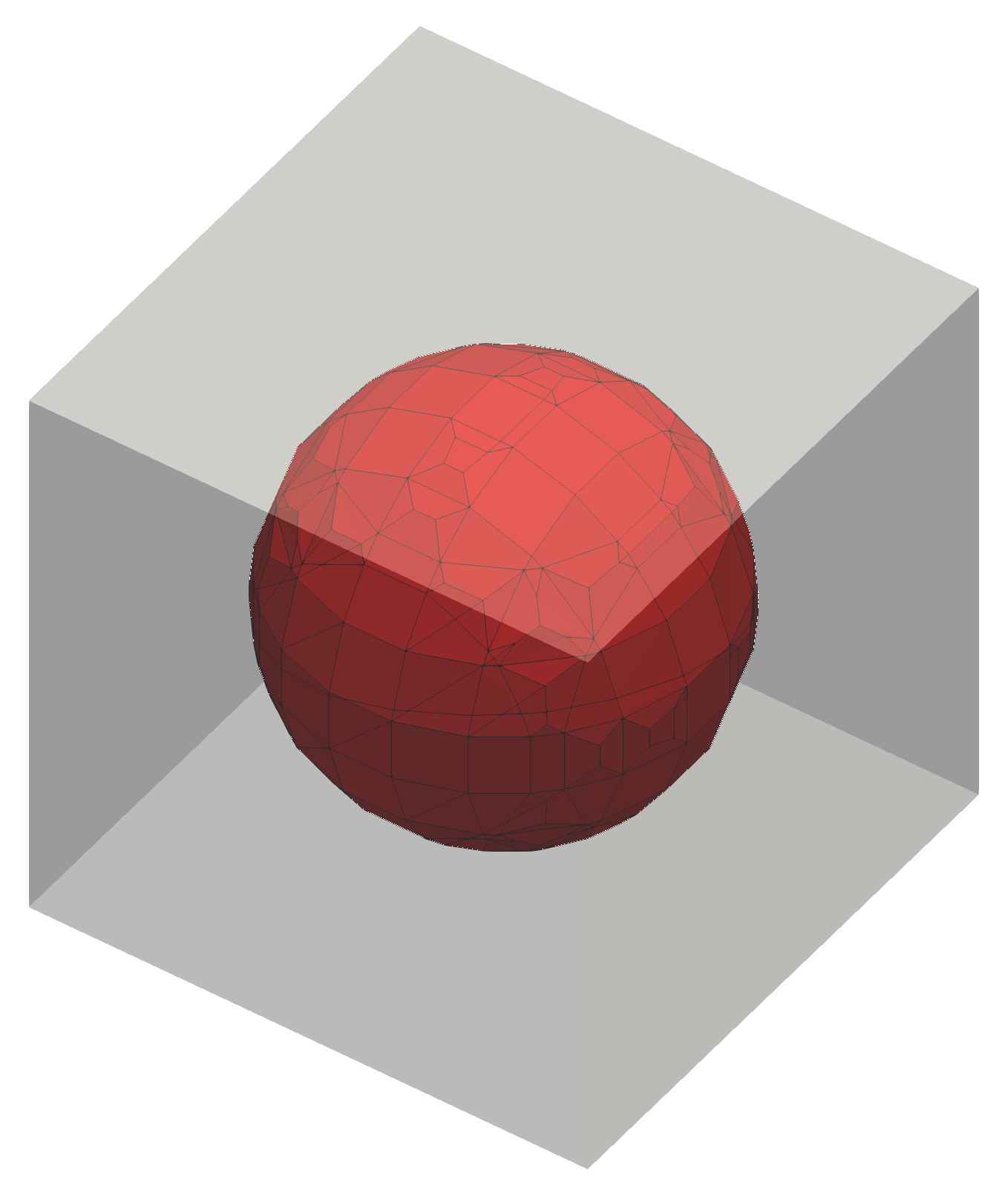}
        \caption{First enrichment surface (shown in red)\label{fig:ver_curved_surface_1}}
    \end{subfigure} %
    ~
    \begin{subfigure}[t]{0.4\textwidth}
        \centering
        \includegraphics[height=1.8in]{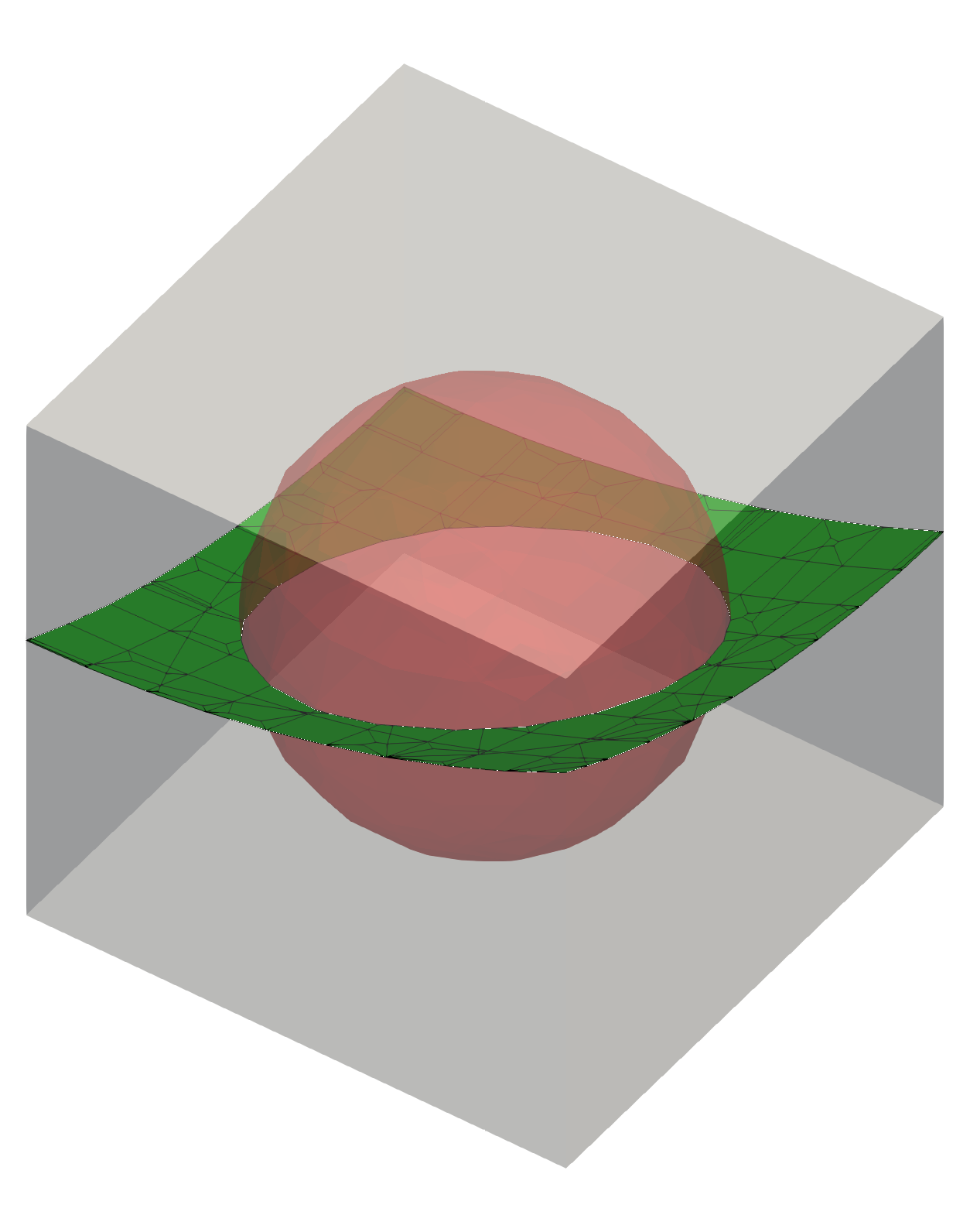}
        \caption{Second enrichment surface (shown in green)\label{fig:ver_curved_surface_2}}
    \end{subfigure}
    
    \begin{subfigure}[t]{0.4\textwidth}
        \centering
        \includegraphics[height=2.2in]{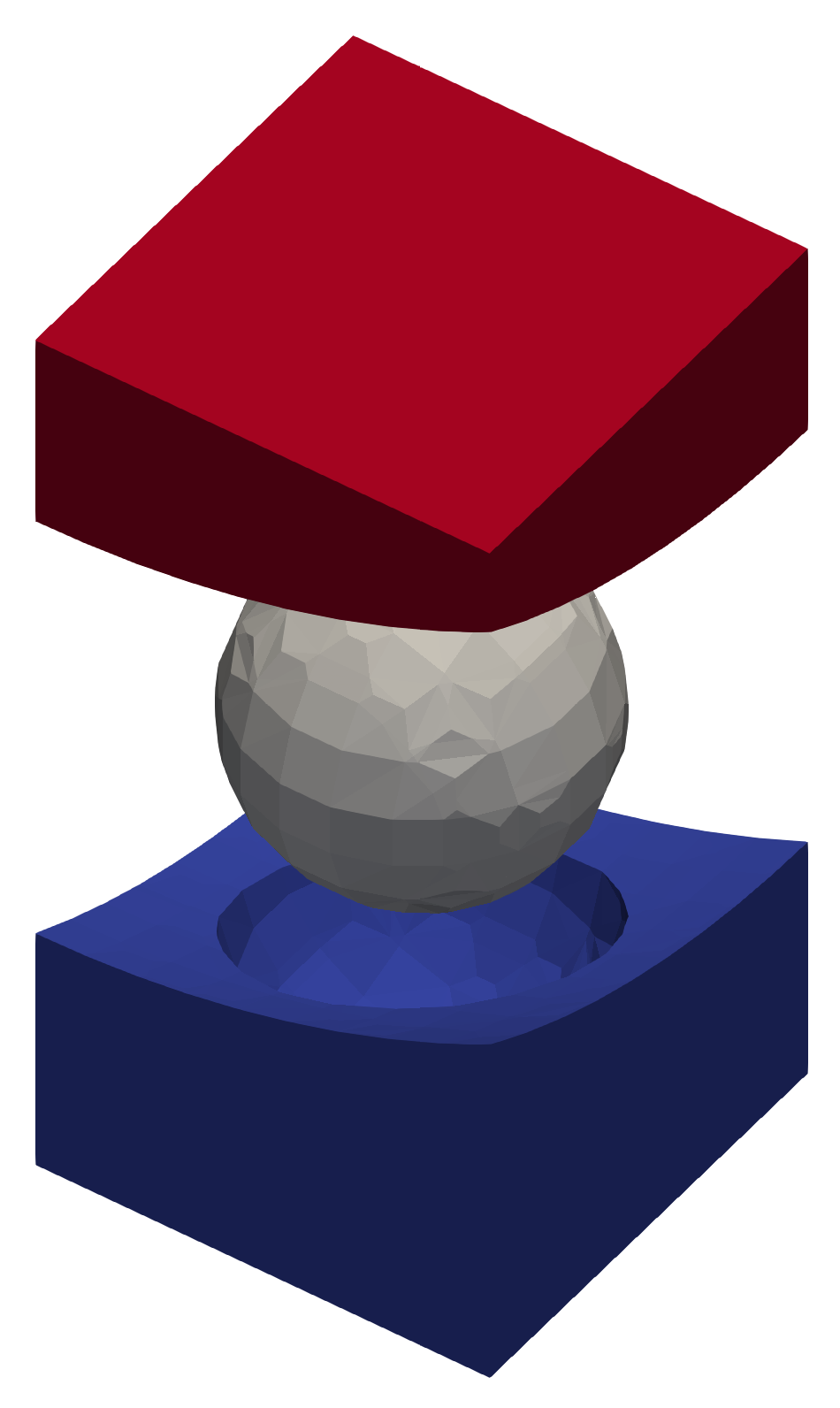}
        \caption{Deformed configuration\label{fig:ver_curved_d}}
    \end{subfigure}
\caption{Enrichment surfaces and results for the verification case of intersecting curved enrichment surfaces with a mixed-element mesh.\label{fig:ver_curved}}
\end{figure}

First, we insert a surface to form a spherical manifold centered in the grid with a radius of $0.37L$, as shown in Figure~\ref{fig:ver_curved_surface_1}.  Since the surface is traction-free, it disconnects the material inside the sphere from the material outside the sphere.  Second, we insert a surface corresponding to a sphere with center $(0,\ 0,\ 2.71L)$, a radius of $2.37L$, and terminating where it intersects the first surface and the boundary of the grid, as shown in Figure~\ref{fig:ver_curved_surface_2}.

To fully test our implementation, we prescribe boundary conditions to distinctly separate each disconnected region of the domain, i.e.,
\begin{align}
u_1\left(0,x_2,x_3\right)=u_2\left(x_1,0,x_3\right) &= u_3\left(x_1,x_2,0\right)=0,\\
u_3\left(0.5L,\, 0.5L,\, 0.5L\right) &= 0.5L,\ \mathrm{and}\\
u_3\left(x_1,x_2,L\right) &= L.
\end{align}

An implementation error would likely result in a nonzero displacement gradient within each disconnected piece of the domain.  Figure~\ref{fig:ver_curved_d} shows the deformed configuration with contours based on the magnitude of the displacement vector, $\left|\bar{u}\right|$.  As expected from a correct implementation, $\left|\bar{u}\right|$ is constant in each of the three pieces of the domain.  Furthermore, the bottom piece correctly experiences no deformation, the sphere correctly translates along $x_3$ by $0.5L$, and the top piece correctly translates along $x_3$ by $L$.  This case also illustrates the ability of our method to evolve enrichment surfaces through a mixed-element mesh.

\subsection{Intersection of Enrichment Surface Fronts}\label{verification_2}

\begin{figure}[ht!]
\centering
    \begin{subfigure}[t]{0.4\textwidth}
        \centering
        \includegraphics[height=1.8in]{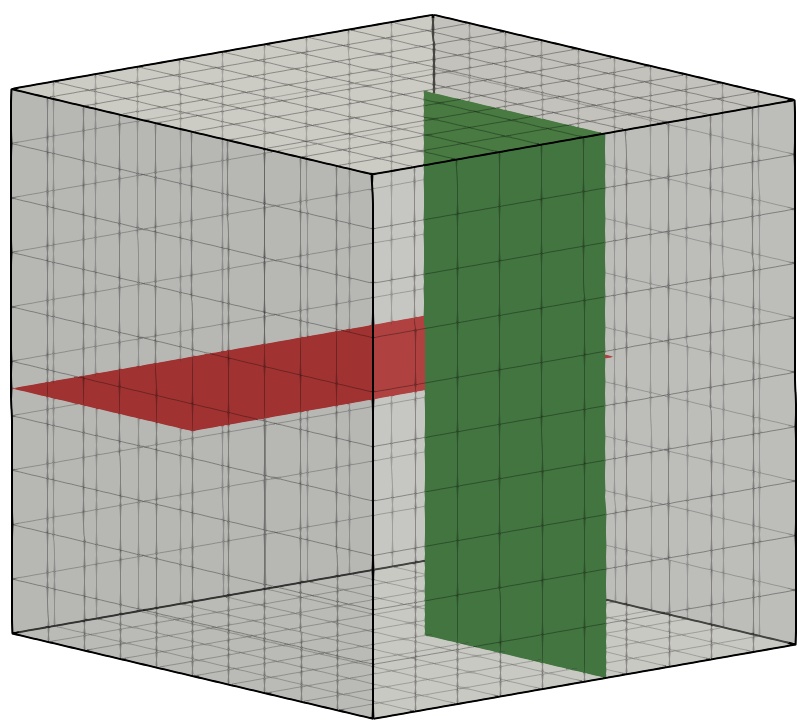}
        \caption{Both enrichment surfaces just before intersection of the fronts\label{fig:ver_fronts_a}}
    \end{subfigure} %
    ~ 
    \begin{subfigure}[t]{0.4\textwidth}
        \centering
        \includegraphics[height=1.8in]{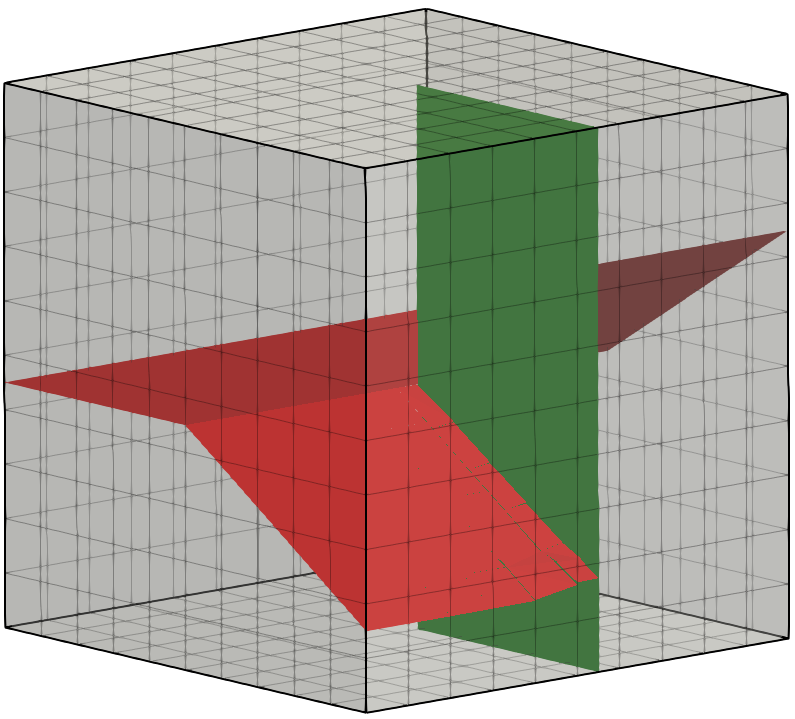}
        \caption{After growth of the red enrichment surface\label{fig:ver_fronts_b}}
    \end{subfigure}

    \begin{subfigure}[t]{0.4\textwidth}
        \centering
        \includegraphics[height=1.8in]{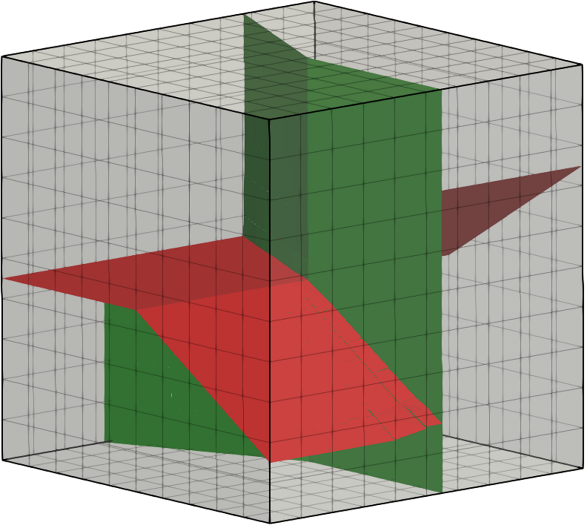}
        \caption{After growth of the green enrichment surface\label{fig:ver_fronts_c}}
    \end{subfigure} %
    ~ 
    \begin{subfigure}[t]{0.4\textwidth}
        \centering
        \includegraphics[height=1.8in]{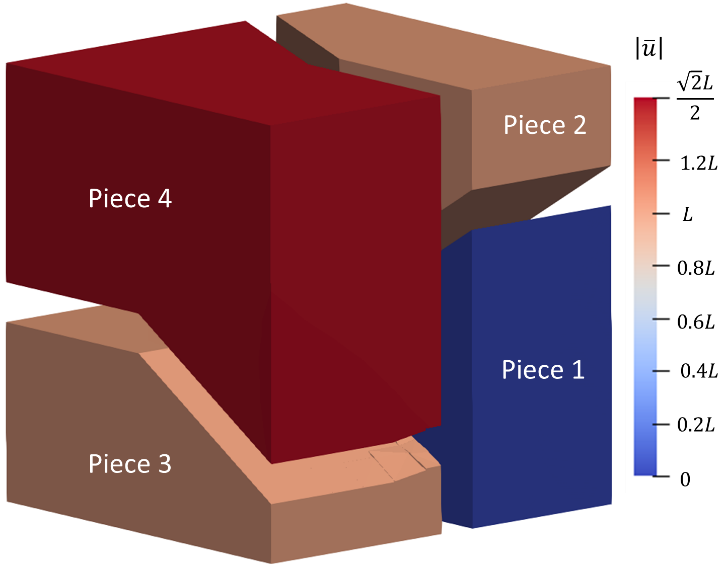}
        \caption{Deformed configuration\label{fig:ver_fronts_d}}
    \end{subfigure}
\caption{Uniform grid, enrichment surfaces, and results for the verification case of intersecting surface fronts.\label{fig:ver_fronts}}
\end{figure}

In this verification case, we consider two enrichment surfaces whose fronts intersect within a uniform grid with side length, $L$.  First, we insert and grow two enrichment surfaces up to the point where the surfaces intersect (Figure~\ref{fig:ver_fronts_a}).  The first enrichment surface (red) now lies in the $x_3=0.45L$ plane, while the second enrichment surface (green) lies in the $x_1=0.45L$ plane.  Next, we grow the red enrichment surface with the normal vector, ${\hat{n}}_1$, specified by  
\begin{align}
{\hat{n}}_1\left(\bar{x}\right)=\begin{cases}
-0.6\hat{j}+\hat{k}, & \mathrm{if}\,  x_1<0.45L, \\
\phantom{+} 0.6\hat{j}+\hat{k}, & \mathrm{otherwise},
\end{cases}
\end{align}
where $\hat{i}$, $\hat{j}$, and $\hat{k}$ are unit vectors along the $x_1$, $x_2$, and $x_3$ axes, respectively.  Figure~\ref{fig:ver_fronts_b} shows enrichment surfaces after the red surface has grown to the boundary of the grid.  Next, we grow the green enrichment surface with normal vector, ${\hat{n}}_2$, specified by
\begin{eqnarray}
{\hat{n}}_2\left(\bar{x}\right)=\begin{cases}
    \hat{i}+0.45\hat{j}, & \mathrm{if}\, x_2<0.45L, \\
    \hat{i}-0.45\hat{j}, & \mathrm{otherwise}.
\end{cases}
\end{eqnarray}

Figure~\ref{fig:ver_fronts_c} shows enrichment surfaces after the green surface has grown to the boundary of the grid.  Finally, we prescribe the following boundary conditions:
\begin{align}
u_1\left(0,x_2,x_3\right) &= u_2\left(x_1,0,x_3\right)=u_3\left(x_1,x_2,0\right)=0,\\
u_1\left(L,x_2,x_3\right) &= L,\ \mathrm{and}\\
u_3\left(x_1,x_2,L\right) &= L,
\end{align}
where the same global coordinate system shown in Figure~\ref{fig:ver_curved_a} is used.  Figure~\ref{fig:ver_fronts_d} shows the deformed configuration with contours based on the magnitude of the displacement vector, $\left|\bar{u}\right|$. As expected, $\left|\bar{u}\right|$ is constant in each piece of the domain, with Piece 1 experiencing no deformation, Piece 2 translating along $x_3$ by L, Piece 3 translating along $x_1$ by L, and Piece 4 translating by $L\hat{i}+L\hat{k}$.  Both of these simulations show that our implementation properly supports the operations needed to enable complex growth of enrichment fronts in 3D.

\clearpage

\section{Comparison of XFEM Fracture Models in the Literature}\label{comparison}

To understand how this work compares to other XFEM approaches for progressive fracture, this section provides enrichment diagrams for several approaches in the literature, illustrating how they would fit within the HHE framework.  The enrichment diagrams developed in Section~\ref{diagrams} provide a concise, clear way of communicating some of the key similarities and differences between methods. They do not communicate enough information to fully characterize a given XFEM approach as many methods differ at a level of detail not revealed by the diagrams.  However, the diagrams do illustrate compatibilities or incompatibilities that exist for a given method, which is often difficult to discern in the literature.

For a comparison of the methods, consider a crack orthogonally intersecting a bi-material interface, which can delaminate via a cohesive connection.  Although the methods extend to 3D, the comparison  will use a 2D situation for simplicity.  The figures below show the material above the bi-material interface in light blue and below in green.  As some methods cannot support an embedded bi-material interface, the bi-material interface will lie on an element boundary when necessary.  The comparison of approaches pays special attention to the compatibilities of fields across enriched elements and the type of cohesive connections between elements on each side of the bi-material interface.  There is a strong interaction between the opening of the embedded crack and the delamination of the bi-material interface.  Fang et al. showed that an enriched cohesive zone is necessary to properly capture this interaction, although they used the term “augmented cohesive zone” since the work was within the context of the augmented finite element method \cite{Fang2011augmented}.  Additionally, Chen et al. highlighted the same issue and arrived at the same conclusion \cite{Chen2014floating}.  Thus, methods that support an enriched cohesive zone across the bi-material interface will yield a more accurate solution, although each of these methods are useful within the proper context.

\begin{figure}[ht!]
\centering
    \begin{subfigure}[t]{0.25\textwidth}
        \centering
        \includegraphics[height=1.5in]{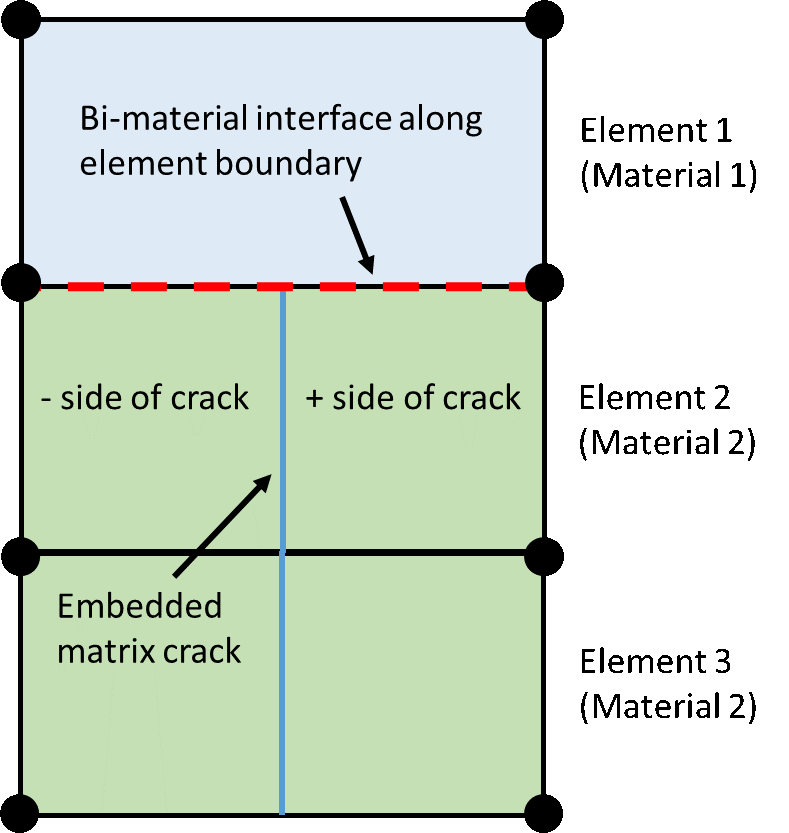}
        \caption{Illustration of mesh and cracks\label{fig:rxfem_a}}
    \end{subfigure} %
    ~ 
    \begin{subfigure}[t]{0.7\textwidth}
        \centering
        \includegraphics[height=1.5in]{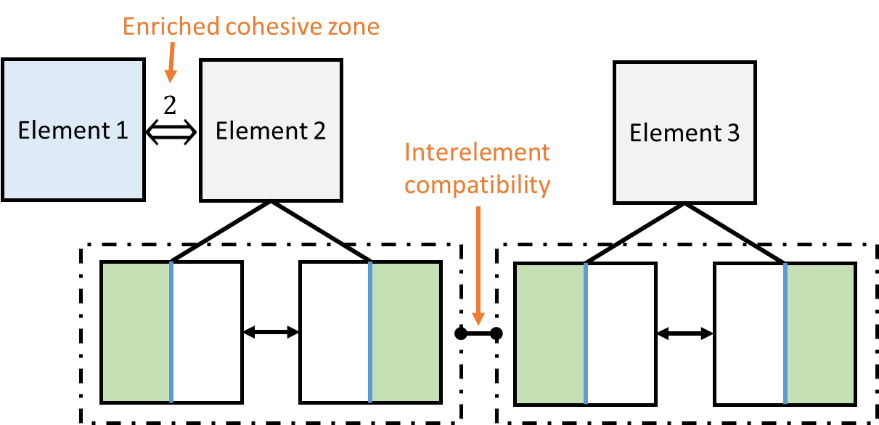}
        \caption{Enrichment diagram\label{fig:rxfem_b}}
    \end{subfigure}
\caption{Enrichment diagram of a crack intersecting a bi-material interface for RXFEM.\label{fig:rxfem}}
\end{figure}

First, consider the regularized-XFEM (RXFEM) method developed by Iarve et al. \cite{Iarve2011Mesh-independent,Iarve2003Mesh}.  For this approach, the bi-material interface must lie along element boundaries to allow intersections between cracks and bi-material interfaces.  Figure~\ref{fig:rxfem_a} illustrates the locations of elements (solid black lines), the bi-material interface (dashed red line), and the embedded crack (solid blue line).  Figure~\ref{fig:rxfem_b} shows the corresponding enrichment diagram for RXFEM, though regularization of the crack is an important detail of RXFEM not captured by these diagrams.  Note that RXFEM maintains compatibility of the fields across elements 2 and 3, which is physically correct.  Additionally, RXFEM correctly uses an enriched cohesive connection between elements 1 and 2, across the bi-material interface.  This results in a cohesive law that independently governs the opening, firstly, between element 1 and the field on the negative side of the crack in element 2 and, secondly, between element 1 and the field on the positive side of the crack in element 2.  RXFEM correctly maintains compatibilities and captures the interaction between the embedded crack and delamination of the bi-material interface; additionally, it is implemented for 3D in BSAM, through a joint effort of AFRL, UDRI, and UTARI \cite{Iarve2011Mesh-independent,Swindeman2013Strength}.  However, the approach currently requires cohesive bi-material interfaces to lie along element boundaries and does not allow cracks to intersect.

\begin{figure}[ht!]
\centering
    \begin{subfigure}[t]{0.25\textwidth}
        \centering
        \includegraphics[height=1.7in]{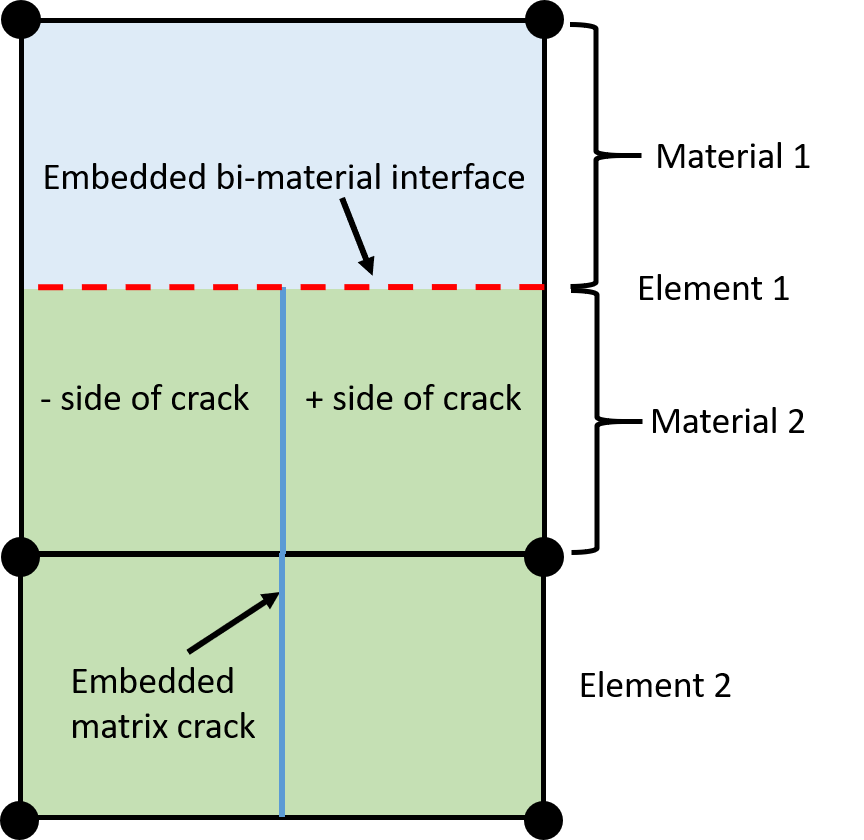}
        \caption{Illustration of mesh and cracks\label{fig:node_meth_a}}
    \end{subfigure} %
    ~ 
    \begin{subfigure}[t]{0.7\textwidth}
        \centering
        \includegraphics[height=1.7in]{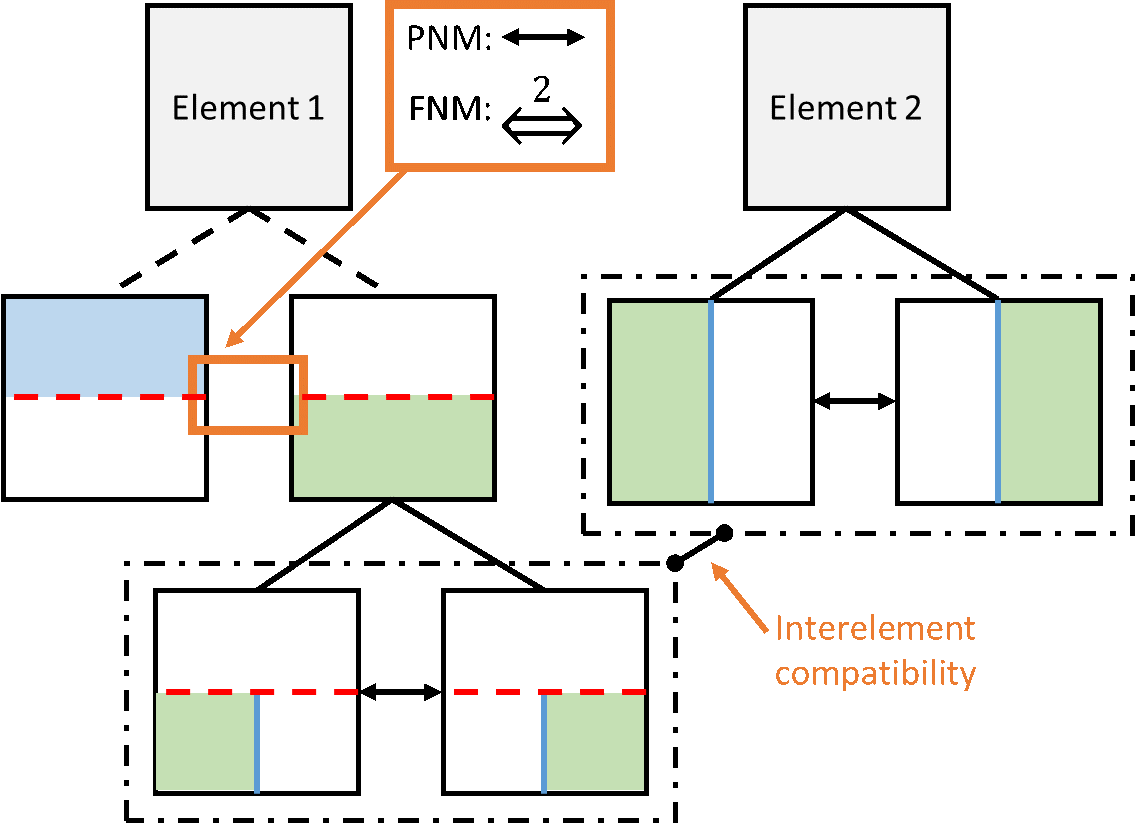}
        \caption{Enrichment diagram\label{fig:node_meth_b}}
    \end{subfigure}
\caption{Enrichment diagram of a crack intersecting a bi-material interface for the phantom-node method (PNM) and floating-node method (FNM).\label{fig:node_meth}}
\end{figure}

Figure~\ref{fig:node_meth} shows the case of a bi-material interface embedded within an element and the corresponding enrichment diagrams for both the phantom-node method (PNM) \cite{Song2006method} and floating-node method (FNM) \cite{Chen2014EXTENED,Chen2014floating}.  As shown in the enrichment diagram, PNM maintains compatibility of the connected fields across elements 1 and 2, which is physically correct.  However, PNM does not use an enriched cohesive connection across the bi-material interface in element 1, which effectively uses the aggregate field for the cohesive connection.  Consequently, PNM would not capture the interaction between the crack opening and delamination of the bi-material interface.

Next, consider FNM, which has also been referred to as the extended phantom-node method.\cite{Chen2014EXTENED,Chen2014floating}  Like PNM, FNM correctly maintains compatibility of connected fields across elements, but unlike PNM, FNM uses an enriched cohesive zone to govern the delamination of the bi-material interface.  FNM has been extended to 3D but restricts the number of cracks that can enter a single element to one.\cite{Chen2017Modelling}  The enrichment diagram does not show some other differences between FNM and PNM, such as the method used to integrate over the physical domain of each field and where the “phantom” or “floating” nodes are located.  PNM places “phantom” nodes at the original locations of nodes in the element that lie on the opposite side of the enrichment for a given field.  However, FNM places “floating” nodes where the enrichment intersects the element boundaries.  Chen et al. discussed this difference in detail. \cite{Chen2014floating} 

\begin{figure}[ht!]
    \centering
    \begin{subfigure}[t]{0.25\textwidth}
        \centering
        \includegraphics[height=1.7in]{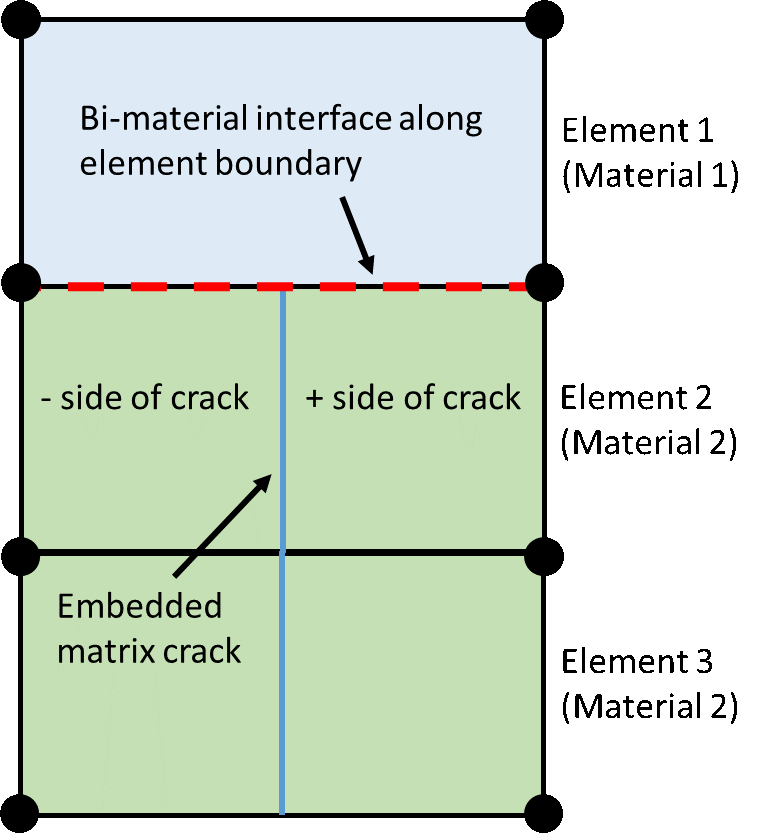}
        \caption{Illustration of mesh and cracks\label{fig:afem_a}}
    \end{subfigure} %
    ~ 
    \begin{subfigure}[t]{0.7\textwidth}
        \centering
        \includegraphics[height=1.7in]{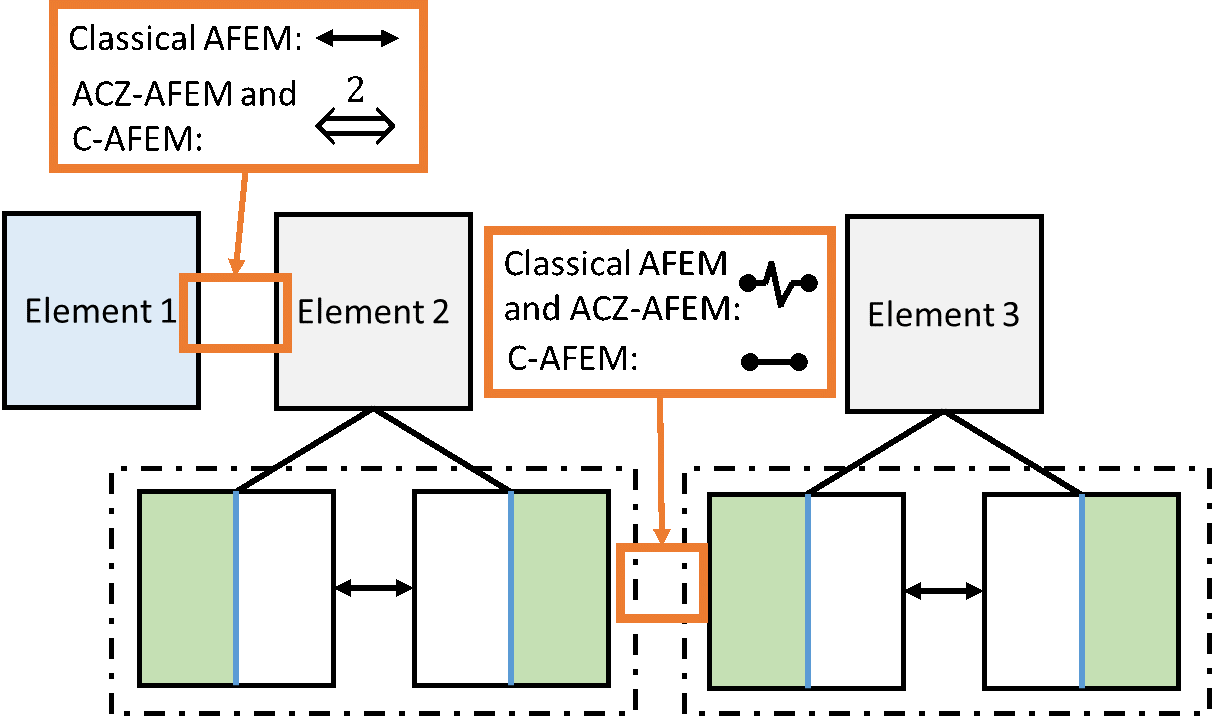}
        \caption{Enrichment diagram for the variants of AFEM\label{fig:afem_b}}
    \end{subfigure}
\caption{Enrichment diagram of a crack intersecting a bi-material interface for augmented finite element methods.\label{fig:afem}}
\end{figure}

Finally, consider the collection of approaches derived from the augmented finite element method (AFEM), which fall into three categories: classical AFEM, AFEM with augmented cohesive zones (ACZ-AFEM), and conforming-AFEM (C-AFEM).  Implementations of AFEM in the literature have focused on embedding discontinuities and have required bi-material interfaces to lie along element boundaries, but in theory, it would not require much effort to alleviate this restriction.  Additionally, all AFEM implementations have been for 3D.  Figure~\ref{fig:afem_a} illustrates the locations of elements, the bi-material interface, which lies between elements 1 and 2, and the crack, which is embedded in elements 2 and 3.  For the situation depicted in Figure~\ref{fig:afem_a}, Figure~\ref{fig:afem_b} shows the corresponding enrichment diagrams for these AFEM models.  Classical AFEM was developed based on Hansbo and Hansbo and PNM to account for the existence of embedded discontinuities that do not conform to the mesh, but the major modification was statically condensing additional degrees of freedom\cite{Ling2009augmented}.  In this sense, AFEM may not fall into the genre of extrinsic XFEM.  However, the element level static condensation introduces interelement incompatibilities.  In the example shown, this strategy results in the field on the negative side of the crack not being continuous across the boundary of elements 2 and 3, which is not physical, as shown in Figure~\ref{fig:afem_b}.  Similarly, the field on the positive side of the crack is not continuous across the same element boundary.  This nonphysical artifact of the method will result in unrealistic stresses near the boundary between elements 2 and 3.  Additionally, classical AFEM employs a standard cohesive zone between elements 1 and 2, which misses the proper coupling between the cohesive zone along the crack and along the interface.  To capture the coupling between intersecting cohesive zones, Fang et al. developed the augmented cohesive zone, which this work generally refers to as an enriched cohesive zone, and the corresponding extension of AFEM (ACZ-AFEM)\cite{Fang2011augmented}.  Finally, very recently, Ma et al. amended the incompatibility introduced across elements in AFEM in what is called conforming-AFEM (C-AFEM)\cite{Ma2019Conforming}.  To maintain compatibility across elements, Ma et al. proposed solving a PDE at the level of an enrichment surface and then solving the global PDE, similar to a local-global approach.  This strategy still avoids the need to introduce additional degrees of freedom into the system, which can reduce the computational cost compared to extrinsic XFEM approaches, but a distributed implementation becomes more complex.  Figure~\ref{fig:afem_b} shows the corresponding enrichment diagram for all three methods.

\section{Discussion and Conclusions}\label{conclusion}

In this article, we presented an extrinsic, continuous-Galerkin extended finite element method for accounting for discontinuities in the solution field.  Like the CG-XFEM methods developed by Hansbo and Hansbo \cite{Hansbo2004finite} and Iarve \cite{Iarve2003Mesh,Iarve2011Mesh-independent}, we account for discontinuities within elements by introducing a new set of DoFs for the enriched elements.  However, we generalize the method proposed by Hansbo and Hansbo to support an arbitrary number of Heaviside enrichments within a single element in a hierarchical fashion without introducing artifacts as seen in the Phantom-Node Method and Augmented Finite Element Method.  This hierarchical view naturally allows for representing element-wise enrichments as a binary tree and enables visual description via enrichment diagrams.  The purpose of this work was to lay the groundwork for a CG-XFEM approach that can accommodate evolving, complex enrichment surfaces, which is a significant step towards a method for modeling progressive fracture.  Towards this end, we made the following novel contributions to the community:
\begin{enumerate}[noitemsep]
    \item We carefully crafted a set of terminology and a lexicon for enrichment diagrams to describe hierarchical Heaviside enrichment (HHE) that clarifies the nuanced concepts within HHE.
    \item We introduced a method to construct a basal field compatibility graph and an algorithm to incrementally update the graph as enrichment surfaces evolve.
    \item We developed a DoF enumeration algorithm that naturally maintains a CG solution across elements and allows cheap updates to the enumeration based on the information in the basal field compatibility graph.
    \item We described the structure of enrichments in an element using enrichment diagrams for several existing extrinsic, XFEM methods in the literature to highlight several subtle differences between the various methods.
\end{enumerate}

We described a derivation of a finite element model for HHE, a hybrid implicit/explicit representation of enrichments, an algorithm to evolve enrichment surfaces, a numerical integration method suitable for volume integrals, an algorithm for constructing a basal field compatibility graph, and a degree-of-freedom enumeration algorithm.  Finally, we provided two verification problems to illustrate that the method allows enrichment surfaces to separate as expected for two situations that are challenging for CG-XFEM, including a case demonstrating that the method supports unstructured, mixed-element meshes.  

Since this work aims to lay the foundation for a progressive fracture model, we restricted the discussion in this paper to modeling open cracks within the context of static 3D linear elasticity.  We intend to follow this work with a description of how cohesive crack models may be implemented within our proposed framework and how enrichment surfaces may be used to model bi-material interfaces in heterogeneous materials.  Subsequently, we plan to apply the framework to modeling complex materials for which it is often difficult to obtain conforming meshes, such as 3D textile composites.  Additionally, we intend to simulate the progressive failure of other composites that are known to develop complex networks of cracks, such as ceramic-matrix composites.  Our algorithms inherently preserve locality, making them ideally suited to both distributed and shared memory parallelization.  We expect to present the details of such a parallelization soon, which will be needed to tackle problems involving progressive fracture in advanced composite materials.  Though not described in this paper, our strategy synergizes well with ongoing work in matrix-free methods for linear systems arising from FEM discretizations, enabling very efficient updates of the element contributions to the system of equations as cracks evolve.\cite{Sundar2007Low-constant,Sampath2008Dendro}

\section*{Acknowledgements}\label{acknowledgements}

The authors would like to acknowledge Professor Endel Iarve at the University of Texas at Arlington, Dr. Michael Braginsky and Dr. Eric Zhou at the University of Dayton Research Institute (UDRI), and Dr. David Mollenhauer at the Air Force Research Laboratory (AFRL) for the underlying concepts of RXFEM that inspired this work.  The authors also would like to thank Dr. Craig Przybyla at AFRL for his advice and Dan Rapking at UDRI for many discussions during the development of this work.  Finally, the authors would like to thank Professor Kurt Maute at the University of Colorado Boulder for comments on a preliminary version of this paper.

\subsection*{Financial disclosure}

The authors would like to acknowledge the support of the Air Force Research Laboratory through contract FA8650-17-C-5269.  Finally, this research was performed in part while M. Keith Ballard held a National Research Council Research Associateship award at the Materials and Manufacturing Directorate of the Air Force Research Laboratory.

\subsection*{Conflict of interest}

The authors declare no potential conflict of interests.

\nocite{*}% Show all bib entries - both cited and uncited; comment this line to view only cited bib entries;
\bibliographystyle{elsarticle-num}
\bibliography{xfem_manuscript}%

\end{document}